\documentclass[twocolumn,secnumarabic,amssymb, nobibnotes, aps, prb]{revtex4-1}
\usepackage{graphicx}
\usepackage{epsfig}
\usepackage{amsmath,amssymb,amsfonts,phonetic}
\usepackage{psfrag}
\usepackage{enumitem}
\usepackage{color}
\usepackage[colorlinks]{hyperref}
\usepackage{epstopdf}
\usepackage{tikz}

\begin{document}

\title{\textbf{ Dynamic transition of vortices into phase slips and generation of vortex-antivortex pairs in thin film Josephson junctions under dc and ac currents. } }

\author{Ahmad Sheikhzada}
\email{asheikhz@odu.edu}
\author{Alex Gurevich}
\email{gurevich@odu.edu}

\affiliation{Department of Physics and Center 
for Accelerator Science, Old Dominion University, Norfolk, VA 23529, USA}


\begin{abstract}

We present theoretical  and numerical investigations of vortices driven by strong dc and ac currents in long Josephson junctions described by a nonlinear integro-differential equation which takes into account nonlocal electrodynamics of films, vortex bremsstrahlung and Cherenkov radiation amplified by the attraction of vortices to the edges of the junction. The work focuses on the dynamics of vortices in Josephson junctions in thin films where the effects of Josephson nonlocality dominate but London screening is negligible. We obtained an exact solution for a vortex driven by an arbitrary time-dependent current in an overdamped junction where the vortex turns into a phase slip if the length of the junction is shorter than a critical length which depends on current. Our analytical and numerical results show that the dynamic behavior of vortices depends crucially on the ohmic damping parameter. In overdamped junctions vortices expand as they move faster and turn into phase slips as current increases. In underdamped junctions vortices entering from the edges produce Cherenkov radiation  generating cascades of expanding vortex-antivortex pairs, which ultimately drive the entire junction into a resistive phase slip state. Simulations revealed a variety of complex dynamic states of vortices under dc and ac currents which can manifest themselves in hysteretic current-voltage characteristics with jumps and regions with negative differential resistance resulting from transitions from oscillating to ballistic propagation of vortices, their interaction with pinning centers and standing nonlinear waves in the junction.

\end{abstract}

\maketitle

\section{Introduction}
\label{sec:intro}

Dynamics of Josephson vortices under strong dc and ac currents \cite{BP,KL,MT} and applications in flux flow oscillators\cite{ffo1,ffo2,ffo3}, multilayer THz radiation sources\cite{thz1,thz2}, or nanoscale superconducting structures for digital memory and quantum computing \cite{qc,jm} have been an area of active investigation, both experimentally and theoretically. Electrodynamics of Josephson vortices has attracted much  attention after the discoveries of high-$T_c$ cuprates and iron-based superconductors in which grain boundaries between misoriented crystallites behave as long Josephson junctions which subdivide the materials into weakly coupled superconducting regions \cite{HM,D}. The latter gives rise to the electromagnetic granularity which is one of the essential obstacles for applications of cuprate and iron-based superconductors \cite{D,AGR}.  Grain boundaries also become performance-limiting defects in superconducting resonator cavities \cite{HP} and thin film multilayer screening structures \cite{AGmlt} for particle accelerators where the amplitudes of the radio-frequency Meissner screening current densities $J(x,t)$ can approach the depairing limit $J_d$. In this case strongly-coupled grain boundaries in Nb or Nb$_3$Sn can behave as long Josephson junctions, even though they may not manifest themselves as weak links in conventional dc magnetization or transport properties of superconductors at much smaller current densities required for 
depinning of vortices.

A conventional theory of Josephson (J) vortices is based on the generic sine-Gordon equation \cite{BP,KL,MT} which is applicable if the phase difference $\theta(x)$ along the junction varies slowly over the magnetic penetration depth. For bulk long junctions, this condition requires small tunneling critical current densities $J_c\ll J_d/\kappa$,  where $\kappa$ is the Ginzburg-Landau parameter \cite{AG}. This condition does not allow using the sine-Gordon approach for high-$J_c$ junctions (such as low-angle grain boundaries) in cuprates and pnictides with $\kappa\simeq 10^2$, and particularly for edge Josephson junctions in thin films where a stray magnetic field $H(x,y)$ outside the junction varies over the Pearl length \cite{pearl} $\Lambda=2\lambda^2/s$ which can be much larger than the London penetration depth $\lambda$ if the film thickness $s$ is much smaller than $\lambda$. For these cases the relation between $\theta(x)$ and $H(x,y)$ becomes nonlocal \cite{KL,AG,IV,agac,mints,MS,kuz,silin,kog,AG3,alf,Abdul}, resulting in mixed Abrikosov-Josephson (AJ) vortices \cite{AG} in which superconducting currents extending over the length $\sim\Lambda$ circulate around a Josephson core of length $l\simeq \xi J_d/J_c$ along the junction, where $l$ is larger than the coherence length $\xi$. Such AJ vortices in which the order parameter in the core is not suppressed have been revealed by transport measurements on low-angle grain boundaries in cuprates \cite{ajgb1,ajgb2,ajgb3}, annular Josephson junctions \cite{ustinov,gol,Ustin,abdul2}, magnetization of thin films \cite{krasnov}, and most recently by STM imaging of step edge junctions in Pb and In atomic monolayers on Si substrates \cite{brun,yosh,rod,2D}. 

Dynamics of J vortices described by the sine-Gordon equation has been investigated in great detail \cite{BP,KL}, but the effects of electromagnetic nonlocality on the properties of vortices in Josephson junctions have been addressed to a much lesser extent. Exact solutions which describe single and periodic AJ vortices driven by strong ac currents in overdamped junctions have been obtained \cite{AG,agac,silin,AG3,alf} yet the nonlinear dynamics of fast vortices in the presence of weak ohmic drag is not well understood. However, it is the behavior of fast vortices in underdamped junctions which becomes markedly different from the conventional sine-Gordon dynamics, because vortices moving with a constant velocity emit Cherenkov radiation due to the fundamental nonlocality of Josephson electrodynamics \cite{MS,Abdul,Kl}. Recent simulations of vortices in long underdamped junctions have shown that the nonlocality can manifest itself in a striking instability of a moving vortex which generates a cascade of expanding vortex-antivortex (V-AV) pairs above a threshold velocity even in nominally low-$J_c$ junctions in which weak nonlocality has been usually disregarded \cite{screp}. This result addresses a broader issue of stability of topological defects driven by external forces and shows that a fast vortex can destroy the global phase coherence in a Josephson junction in a way similar to crack propagation resulting from the pileup of dislocations of opposite polarity \cite{disl}.

Our previous results \cite{screp} obtained for an infinitely long junction bring about the following issues related to the dynamics of vortices in junctions of finite length which are most relevant to experiments: 1. What happens to AJ vortices driven by strong currents in a finite junction where in addition to the Cherenkov radiation, a vortex also radiates as it accelerates and decelerates due to its attraction to the edges of the junction? 2. How can the finite length of the junction affect generation of V-AV pairs by the radiation field of moving vortices?  3. How can the finite size effects change the structure of a static or moving AJ vortex, and whether they could cause a transition from a vortex to a phase slip state in which $\theta(t)$ becomes uniform along the junction?  4. What are  manifestations of Josephson nonlocality in the dynamics of vortices in finite junctions, as compared to J vortices described by the sine-Gordon equation \cite{mac,HLO,srid,physc}?  Addressing these issues is the goal of this work in which we investigate AJ vortices driven by strong currents in thin film junctions.            

The paper is organized as follows. In section \ref{sec:elec} we introduce the main integro-differential equations of nonlocal Josephson electrodynamics (NJE) which describe $\theta(x,t)$ in junctions of different thin film  geometries and specify the conditions under which the nonlocality becomes dominant. These equations were then solved both numerically and analytically in the extreme nonlocal limit. In section \ref{sec:overd} we present an exact solution of NJE equations for AJ vortex driven by an arbitrary time-dependent transport current in overdamped junction of finite length. In section \ref{sec:dc} we present numerical simulations of AJ vortices driven by dc current at different damping constants. A dynamic transition from AJ vortices to phase slip is shown to occur due to expansion of vortex core in overdamped junctions, and due to Cherenkov radiation in underdamped junctions. It turns out that generation of vortex-antivortex pairs in finite junctions can occur at much larger damping constants than in infinite junctions. In section \ref{sec:ac} we present numerical simulations of AJ vortices under ac current. Implications of our results are discussed in section \ref{sec:disc}.

\section{NJE Equations}
\label{sec:elec}
Dynamics of the gauge invariant phase difference $\theta(x,t)$ on a weakly coupled long Josephson junction is described by the sine-Gordon equation\cite{BP,KL,MT}
\begin{equation}
\ddot{\theta}+\eta\dot{\theta}=\lambda_J^2\theta''-\sin\theta+\beta,
\label{sg}
\end{equation}
where prime and overdot denote partial derivatives with respect to coordinate $x$ and dimensionless time $\omega_J t$, $\omega_J=(2\pi c J_c/\phi_0 C)^{1/2}$ is Josephson plasma frequency, $J_c$ is the junction tunneling critical current density, $\phi_0$ is magnetic flux quantum, $C$ is specific capacitance per unit area of the junction, $c$ is the speed of light, $\lambda_J=(c\phi_0/16\pi^2\lambda J_c)^{1/2}$ is Josephson penetration length, $\eta=1/\omega_JRC$ is damping constant due to ohmic quasiparticle resistance $R$,  and $\beta=J/J_c$ is the dimensionless uniform transport current density across the junction. 
Equation (\ref{sg}) implies a local relation between $\theta(x,t)$ and the magnetic field $H(x,t)$ produced by vortex currents, both varying over the same length $\lambda_J$ which is assumed to be much larger than $\lambda$. If this condition is not satisfied, $\theta(x,t)$ and $H(x,t)$ vary over {\it different} length scales, and the relation between $\theta(x,t)$ and $H(x,t)$ becomes nonlocal. In this case the equation for $\theta(x,t)$ in an infinite junction takes the form \cite{AG,IV,agac,mints,MS,kuz,silin,kog,AG3,alf,Abdul}
\begin{gather}
    \ddot{\theta}+\eta\dot{\theta}=\frac{l_0}{\pi}\int_{-\infty}^{\infty}G\left(|x-u|\right)\frac{\partial^2\theta}{\partial u^2}du -\sin\theta+\beta,
    \label{int} \\
    l_0=\frac{\lambda_J^2}{\lambda}=\frac{c\phi_0}{16\pi^2\lambda^2J_c}.
    \label{l0}
\end{gather}
Equation (\ref{int}) describes nonlocal dynamics of $\theta(x,t)$ and $H(x,t)$ varying over any length scale larger than $\xi$. Here the geometry-dependent kernel $G(x,u)$ diverges logarithmically at $x=u$ and decreases with $u$ if $|x-u|$ exceeds the relevant magnetic penetration depth.  For instance, $G(x)=K_0(x/\lambda)$ for a planar junction in a bulk superconductor, where $K_0(x)$ is the modified Bessel function \cite{AG}. For an edge junction in a thin film of thickness $s\ll\lambda$, the kernel is $G(x)=\pi [\textbf{H}_0(x/\Lambda)-Y_0(x/\Lambda)]/2$, where $\Lambda=2\lambda^2/s$, and $\textbf{H}_0(x)$ and $Y_0(x)$ are the Struve and Bessel functions, respectively \cite{mints,kog}. For an overlap junction in a thin film, $G(x) = \ln\coth\left(\pi |x|/4s\right)$ \cite{alf} also diverges logarithmically at $x=0$ but decreases exponentially over the length $2s/\pi$ shorter than $\lambda$ if $s\ll \lambda$.  

If $\theta(u)$ varies slowly over the scale on which $G(x)$ decreases rapidly, $\theta''(u)$ in Eq. (\ref{int}) can be replaced with $\theta''(x)$ and taken out of the integral. In this case Eq. (\ref{int}) reduces to Eq. (\ref{sg}) provided that $\int_{-\infty}^\infty G(x)dx$ converges. The latter is indeed the case for bulk and overlap junctions for which $G(x)$ decreases exponentially at large $x$. For bulk junctions, Eq. (\ref{int}) reduces to Eq. (\ref{sg}) if $\theta(x)$ varies slowly over $\lambda$. However, for an edge junction in a thin film the kernel $G(x)=\pi[\textbf{H}_0(x/\Lambda)-Y_0(x/\Lambda)]/2$ in the limit of $s\to 0$ decreases as $1/x$ due to long-range stray field outside the film at $x>\Lambda$, and the integral $\int_0^\infty G(x)dx$ diverges logarithmically. In this case Eq. (\ref{int}) reduces to Eq. (\ref{sg}) at $J_c\ll J_d\xi/\Lambda$ only if the effect of finite film thickness in $G(x)$ is taken into account \cite{kuz}.  

Generally, solutions of Eq. (\ref{int}) for a vortex traveling with a constant velocity $v$ can only be obtained numerically. Yet in the weak-coupling local limit of $\lambda_J\gg \lambda$ and $\eta\to 0$, Eq. (\ref{int}) reduces to Eq. (\ref{sg}) which has the well-known solution describing  a moving J vortex \cite{BP,KL}:
\begin{equation}
\theta(x,t)=4\tan^{-1}\exp\left[\frac{x-vt}{\lambda_J\sqrt{1-(v/c_s)^2}}\right].
\label{sgkink}
\end{equation}
Here the length of the vortex $L(v)=\lambda_J\sqrt{1-v^2/c_s^2}$ shrinks as it moves faster due to the ``Lorentz-contraction", with the Swihart velocity $c_s=\lambda_J\omega_J$ being 
the maximum speed of phase waves \cite{BP}. Therefore, the sine-Gordon equation becomes inadequate at $\eta\ll 1$ and high vortex velocities 
at which $L(v)\sim \lambda$ or $L(v)\sim \Lambda$ for edge junctions in thin films. 
Numerical simulations of Eq. (\ref{sg}) for a J vortex driven by a dc current at a finite $\eta$ have shown that  
$L(v)$ decreases with $v$ at $\eta < 1$ but increases with $v$ at $\eta>1$ \cite{KL}.  

In the extreme nonlocal limit of $\lambda_J\ll\lambda$ the analytical solution of Eq. (\ref{int}) for a driven AJ vortex in an overdamped long junction with $\eta\gg 1$ is given by \cite{AG}:
\begin{gather}
\theta(x,t)=\pi+\sin^{-1}\beta+2\tan^{-1}[(x-vt)/l(v)], 
\label{ajkink} \\
l(v)=\frac{l_0}{\sqrt{1-\beta^2}}, \qquad v(\beta)=\frac{\beta l}{\tau}.
\label{lv}
\end{gather}
Here the length $l_0$ of the phase core of AJ vortex along the junction is defined by Eq. (\ref{l0}), and $\tau=\eta/\omega_J=\phi_0/2\pi c R J_c$. As follows from Eq. (\ref{lv}), the AJ vortex expands as it moves faster, similar to the behavior of overdamped J vortex. Equations (\ref{sgkink}) and (\ref{ajkink}) describe solitonic $2\pi$ kinks, neither of which produce any radiation wakes behind a moving vortex. For J vortex, the lack of radiation is due to the Lorentz invariance of Eq. (\ref{sg}) at $\eta=0$, whereas the radiation field for AJ vortex at $\eta\gg 1$ is suppressed by strong dissipation.  In the general case which includes  
the electromagnetic nonlocality, ohmic damping and the displacement current in Eq. (\ref{int}), radiation produced by vortices is essential, particularly in finite junctions, as shown below. 
 
 \subsection{Cherenkov radiation and instability}
 
 \begin{figure} 
\includegraphics[width=\columnwidth]{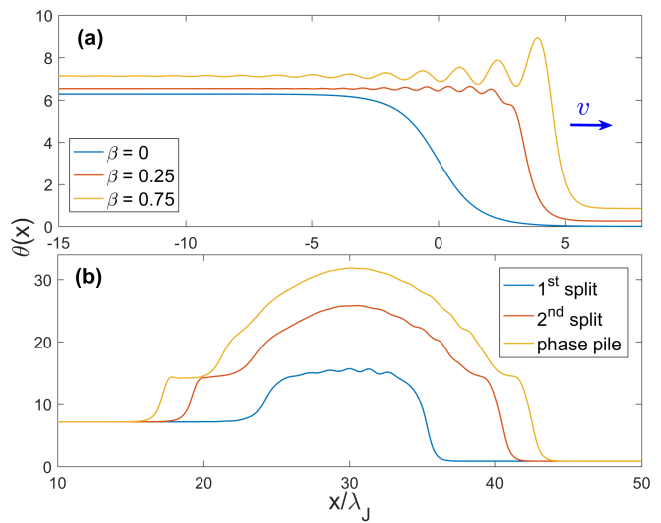}
\caption{\label{scireppro} (a) Wakes of Cherenkov radiation behind a moving vortex in an infinite junction calculated from Eq. (\ref{int}) with $G(x)=K_0(x/\lambda)$ for $\lambda_J/\lambda = 10$, $\eta = 0.07$ and different $\beta$.  (b) Initial stage of vortex instability at $\beta=0.76$.}
\end{figure}

Unlike Eq. (\ref{sg}), the general Eq. (\ref{int}) at $\eta=0$ is not Lorentz-invariant, so a uniformly moving vortex can radiate Cherenkov waves $\delta\theta(x,t) \propto \exp(ikx-i\omega_k t)$ with the phase velocities $\omega_k/k$ smaller than $v$ \cite{MS,Abdul}. Setting $\theta(x,t)=\theta_\infty+\delta\theta(x,t)$ where $\sin\theta_\infty=\beta$, and linearizing Eq. (\ref{int}) with respect to small disturbances $\delta\theta(x,t)$ for a uniform dc current and $\eta=0$, yields the dispersion relation $\omega_k^2=[\cos\theta_\infty+l_0k^2G(k)]\omega_J^2$. Thus, the condition of Cherenkov radiation $kv >\omega_k$ is given by:
\begin{equation}
kv > \omega_J\left[ \sqrt{1-\beta^2}+l_0k^2G(k) \right]^{1/2},
\label{cher}
\end{equation}
where $G(k)$ is the Fourier image of $G(x)$ and $l_0=\lambda_J^2/\lambda$.  Here $G(k)$ decreases as $1/k$ at $k>\Lambda^{-1}$ so Eq. (\ref{cher}) is satisfied if $k>k_c$, where the maximum wavelength $\L_c=2\pi/k_c$ increases with $v$. For a bulk junction, we have $G(k)=\lambda/\sqrt{1+\lambda^2k^2}$, so the threshold $k_c$ at which  Eq. (\ref{cher}) becomes equality can be evaluated in the limit of $\lambda/\lambda_J\ll 1$  by expanding $(1+\lambda^2k_c^2)^{-1/2}\approx 1 - \lambda^2k_c^2/2$ and solving the resulting bi-quadratic equation for $k_c$:
\begin{equation}
k_c^2\lambda^2=1-\frac{v^2}{c_s^2} + \left[ \left(1-\frac{v^2}{c_s^2}\right)^2 + \frac{2\lambda^2}{\lambda_J^2}\sqrt{1-\beta^2} \right]^{1/2}\!\!.
\label{kc}
\end{equation}
The maximum Cherenkov wavelength $\L_c = 2\pi/k_c$ thus increases as $\beta$ and $v$ increase, approaching  
\begin{equation}
\L_{c}\to \frac{2^{3/4}\pi \sqrt{\lambda\lambda_J}}{(1-\beta^2)^{1/8}}, \qquad v\to c_s.
\label{lac}
\end{equation}
Hence $k_c^2\lambda^2\ll 1$, which justifies the above expansion of $G(k)$ in small $k_c$ at $\lambda/\lambda_J\ll 1$.  Equations (\ref{kc})-(\ref{lac}) show that the nonlocality 
of Eq. (\ref{int}) results in Cherenkov radiation behind a uniformly moving J vortex even in a weakly-coupled junction $\lambda\ll\lambda_J$ which is usually described by the sine-Gordon equation (\ref{sg}).  Thus, the approximation of Eq. (\ref{int}) with Eq. (\ref{sg}) can miss essential effects in the dynamics of Josephson vortices.  

These effects are illustrated by Fig. \ref{scireppro} which shows results of numerical simulations of Eq. (\ref{int}) in a bulk underdamped junction biassed by a dc current in a nominally local Josephson limit of 
$\lambda_J=10\lambda$, $\eta=0.07$ and $G(x)=K_0(x/\lambda)$. Yet Eq. (\ref{int}) reveals the effects which are not captured by Eq. (\ref{sg}), particularly a Cherenkov wake behind a uniformly moving J vortex which becomes apparent at $\beta=0.25$ and reaches about $1/3$ of the hight of the $2\pi$ phase kink in J vortex at $\beta=0.75$. Cherenkov radiation can result in a drag force which can be much stronger than the conventional ohmic drag in underdamped junctions \cite{screp}.   

A vortex moving uniformly becomes unstable at $\beta>\beta_i$, the instability develops at the maximum of Cherenkov wake which reaches a 
critical value $\theta_c \approx 8.65-8.84$, depending on $\eta$ and the junction geometry \cite{screp}. Here $\theta_c$ is confined within $5\pi/2 <\theta_c <3\pi$ where 
a uniform state of a Josephson junction is unstable \cite{BP,KL}. As the velocity increases, the wake behind the moving vortex grows and widens and eventually 
becomes unstable due to the appearance of a trailing critical nucleus being in the  
$\pi$-junction state \cite{BP,KL}. In this case Eq. (\ref{int}) has no steady-state vortex solutions at $J > J_i <J_c$  \cite{screp}.

The instability shown in Fig. \ref{scireppro} originates at the maximum of the Cherenkov wake which starts
growing and eventually turning into an expanding V-AV pair. As the size of this pair grows, it generates enough Cherenkov radiation to produce two more V-AV 
pairs which in turn produce new pairs. Continuous generation of V-AV pairs results in an expanding dissipative domain in which vortices accumulate at the right side, antivortices accumulate at the left side, while dissociated vortices and antivortices pass through each other in the middle. As a result, $\theta(x,t)$ evolves into a growing ``phase pile" with the maximum $\theta_m(t)$ increasing 
approximately linear with time and the edges propagating with a speed which can be both smaller and larger than $c_s$, the phase difference $\theta(-\infty)-\theta(\infty)=2\pi$ between the edges remains fixed. The Cherenkov vortex instability and the phase pile dynamic state was obtained by simulations of Eq. (\ref{int}) for different junction geometries and $10^{-3}<\eta<0.53$ \cite{screp}. Evidences of vortex Cherenkov instability were observed in numerical simulations of multilayer annular junctions \cite{Ustin}.

A cascade of expanding V-AV pairs generated by Cherenkov wake suggests that the dynamics of the phase pile state can be affected significantly by the junction length. 
Indeed, a vortex moving in a finite junction gets attracted to its edges which results in deceleration or acceleration of the vortex as it enters or exits the junction. In turn, the 
vortex moving with a time-dependent velocity $v(t)$ produces the Larmor radiation (bremsstrahlung) which adds to the Cherenkov radiation.  Both Cherenkov and Larmor 
contributions produce electromagnetic waves which get reflected from the edges of the junction, forming nonlinear standing waves which affect both dynamics of vortices and the generation of new V-AV pairs. To address these issues, Eq. (\ref{int}) should be generalized to take into account the junction geometry.  
 
 \subsection{Finite junctions in the nonlocal limit}
 
\begin{figure} 
\includegraphics[width=\columnwidth]{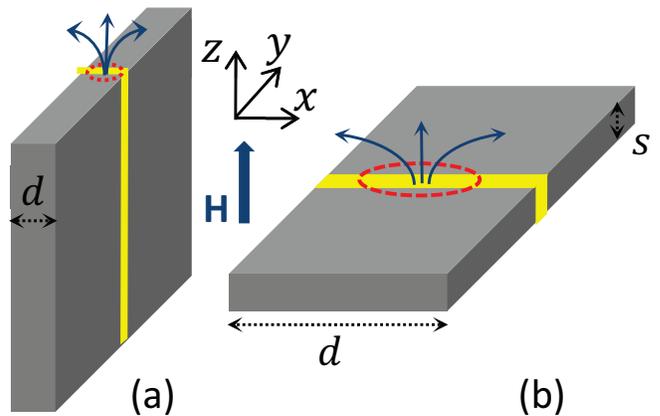}
\caption{\label{junc} Geometries of a Josephson junction in a thin film with the vortex parallel (a) and perpendicular (b) to the broad face of the film.}
\end{figure}

Consider a junction of length $d$ in a film where a vortex is either perpendicular or parallel to the broad surface of the film, 
as shown in Fig. \ref{junc}. Here Fig. \ref{junc} (a) is relevant to a polycrystalline superconducting screen in which the Josephson junction models 
a grain boundary perpendicular to the film, whereas Fig. \ref{junc} (b) represents an edge junction. To derive the equation for $\theta(x,t)$, we start with the superconducting current density:
\begin{gather}
J_x = -\frac{c}{4\pi\lambda^2}\left(\frac{\phi_0}{2\pi}\frac{\partial \varphi}{\partial x}+A_x\right),
\label{jx}\\
J_y = -\frac{c}{4\pi\lambda^2}\left(\frac{\phi_0}{2\pi}\frac{\partial\varphi}{\partial y}+A_y\right),
\label{jy}
\end{gather}
where $\mathbf{A}$ is the vector potential, $\varphi$ is the phase of the order parameter, and $\phi_0=\pi\hbar c/|e|$. The current continuity condition $\partial_xJ_x+\partial_yJ_y=0$ can be satisfied by expressing  
$J_x=\partial_y g$ and $J_y=-\partial_x g$ in terms of a stream function $g(x,y,t)$.
From Eq. (\ref{jx}), it follows that any nonuniform phase difference $\theta(x)=\varphi(x,-0)-\varphi(x,+0)$ on the junction results in a discontinuity of  $J_x(x,+0)-J_x(x,-0)=(c\phi_0/8\pi^2\lambda^2)\partial_x\theta(x)$, and a jump of the normal derivative in the stream function at $y=0$:
\begin{equation}
\frac{\partial g(x,y)}{\partial y}\Big|_{y=+0}-\frac{\partial g(x,y)}{\partial y}\Big|_{y=-0}=\frac{c\phi_0}{8\pi^2\lambda^2}\frac{\partial\theta}{\partial x}.
\label{wbc}
\end{equation}
Excluding $\varphi$ from Eqs. (\ref{jx}) and (\ref{jy}) yields
\begin{equation}
\nabla^2g-\frac{cH}{4\pi\lambda^2}=\frac{c\phi_0}{8\pi^2\lambda^2}\frac{\partial\theta}{\partial x}\delta(y),
\label{lond}
\end{equation}
where $\delta(y)$ provides the boundary condition (\ref{wbc}), and $H=\nabla_z \times\mathrm{\mathbf{A}}$ is the $z$ component of the magnetic field. For a parallel vortex in a 
thin film shown in Fig. \ref{junc} (a), we have $g=cH/4\pi$ and  Eq. (\ref{lond}) yields the London equation for $H(x,y)$. For a perpendicular vortex in an edge junction, $H(x,y)$ in Eq. (\ref{lond}) 
is expressed in terms of $g(x,y)$ using the Biot-Savart law, which turns Eq. (\ref{lond}) into an integro-differential equation. The nonuniform Eq. (\ref{lond}) can be solved using the Green function which is nothing but the solution of the London equation for either a parallel A  vortex \cite{stej} or a perpendicular Pearl vortex \cite{pearl} for the cases shown in Figs. \ref{junc} (a) and (b), respectively.   
A general solution for $g(x,y)$ is rather cumbersome, so we consider simpler cases of a thin film with $d<\lambda$ and a bridge with $d < \Lambda$ for which self-field effects 
and the London screening are inessential. Then $\mathbf{A}$ and $H$ in Eqs. (\ref{jx}), (\ref{jy}) and (\ref{lond}) can be neglected, bias current density $J$ is uniform across the film, and Eq. (\ref{lond}) reduces to the Poisson 
equation for both geometries shown in Fig. \ref{junc}:
\begin{equation}
\nabla^2g=\frac{c\phi_0}{8\pi^2\lambda^2}\frac{\partial\theta}{\partial x}\delta(y).
\label{pu}
\end{equation}
Setting $x=0$ in the middle of the film, we obtain $g(x,y)$ which satisfies the boundary condition $J_x(\pm d/2,0)=\partial_yg(\pm d/2,y)=0$ at the junction edges \cite{stej}:
\begin{gather}
g(x,y)= -Jx-
\nonumber \\
\frac{c\phi_0}{32\pi^3\lambda^2}\int_{-d/2}^{d/2}\ln\frac{\cosh\frac{\pi y}{d}+\cos\frac{\pi}{d}(x+u)}{\cosh\frac{\pi y}{d}-\cos\frac{\pi}{d}(x-u)}\frac{\partial\theta(u)}{\partial u}du.
\label{gxy}
\end{gather}
Using Eq. (\ref{gxy}), the current density $J_y(x)=-\partial_x g(x,0)$ through the junction is calculated. Equating $J_y(x,0)$ to the sum of Josephson, resistive, and 
displacement current densities, and integrating by parts as shown in Appendix A, we obtain the following equation for $\theta(x,t)$:  
\begin{gather}
\ddot{\theta}+\eta\dot{\theta} +\sin\theta-\beta =
\nonumber \\ 
\epsilon\!\int_{-1/2}^{1/2}\!\ln\left|\frac{2}{\sin\pi x-\sin\pi u}\right|\theta''(u)du,
\label{main} \\
\epsilon=\frac{l_0}{\pi d}=\frac{c\phi_0}{16\pi^3\lambda^2dJ_c},
\label{eps}
\end{gather}
where $x$ and $u$ are expressed in units of $d$, and the prime denotes differentiation with respect to the dimensionless coordinate $x$ along the junction. If the geometry-dependent screening effects caused by the vector-potential $\textbf{A}$ in Eqs. (\ref{jx}) and (\ref{jy}) are negligible, $\theta(x,t)$ is described by Eq. (\ref{main}) for both cases  shown in  Fig. \ref{junc}. We will use Eq. (\ref{main}) for the calculations of vortices in Josephson junctions, and the average power $\bar{P}$ dissipated per unit height of the junction: 
\begin{equation}
\bar{P}=\frac{\eta P_0}{T}\int_{0}^{T}dt\int_{-1/2}^{1/2} \dot\theta^2(x,t) dx,
\label{diss}
\end{equation}
where $P_0=\phi_0J_c \omega_J d/2\pi c$. Equations (\ref{int}) and (\ref{main}) take into account only ohmic losses but disregard radiation from a thin film junction into free space. The radiation losses are 
negligible due to a big mismatch of impedances of a superconductor and vacuum \cite{rad1,rad2}, except for the extreme case of underdamped junctions with $\eta\ll 1$. In this paper we calculate dynamics 
of vortices in overdamped and moderately underdamped junctions with $\eta > 0.2$ for which the effect of radiation to free space on the power $\bar{P}$ and $\theta(x,t)$ in Eq. (\ref{main}) is negligible.

\section{Exact solution for a moving overdamped AJ vortex}
\label{sec:overd}
Equation (\ref{main}) with the non-negligible term $\ddot{\theta}$ can only be solved numerically. Yet an exact solution for $\theta(x,t)$ in a vortex 
driven by an arbitrary current $\beta(t)$ in an overdamped junction with $\eta\gg 1$ can be obtained by introducing the dimensionless complex potential:
\begin{equation}
w(z) = \varphi(x,y)+ig(x,y).
\label{w}
\end{equation}
Here $z=x+iy$ are complex coordinates in units of $d$, $w(z)$ and $g(z)$ are in units of $w_0=c\phi_0/8\pi^2\lambda^2$, $\varphi$ is the phase of the order parameter $\Psi(z)=\Delta\exp[i\varphi(x,y)]$, and $\Delta$ is assumed independent of $z$. If $\textbf{A}$ in Eqs. (\ref{jx}) and (\ref{jy}) is negligible, $g(x,y)$ and $\varphi(x,y)$ are related by the Cauchy-Riemann conditions $\partial_x\varphi=\partial_y g$ and $\partial_y\varphi=-\partial_x g$ so that $w(z)$ is an analytic function, and Eqs. (\ref{jx}) and (\ref{jy}) can be written in the dimensionless complex form: 
\begin{equation}
j_x-ij_y=-2\pi\epsilon\frac{dw}{dz},
\label{J}
\end{equation}
where $j_x$ and $j_y$ are in units of $J_c$, and $\epsilon=c\phi_0/16\pi^3\lambda^2dJ_c$ is the same as in Eq. (\ref{eps}). Calculation of $\theta(x,t)$ and $\textbf{J}(x,y)$ then reduces to finding two analytic functions $w_1(z)$ and $w_2(z)$, where $w_1(z)$ has no poles in the upper half-plane $y>0$, and $w_2(z)$ has no poles in the lower half plane $y<0$, so that there are no singularities in the resulting current flow defined by Eq. (\ref{J}). Here $\theta(x)=w_2(x,0)-w_1(x,0)$, and $w_1(z)$ and $w_2(z)$ are linked by continuity of $g(x,y)$ and $j_y(x,y)$ at $y=0$:
\begin{equation}
 \tau\partial_t\theta+\sin\theta= j_y(x,0).
\label{bc}
\end{equation}   

\begin{figure} 
\includegraphics[width=\columnwidth]{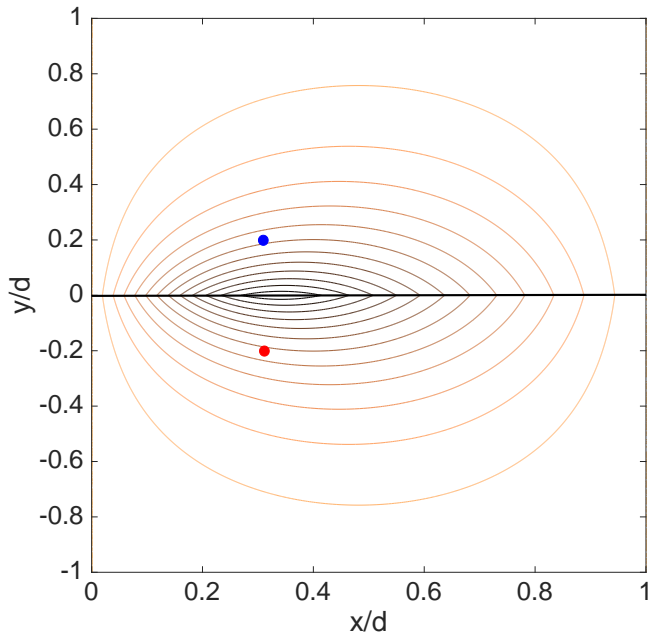}
\caption{\label{ajcur} Current streamlines in the AJ vortex with $u=0.3d$ and $l=0.2d$ calculated from Eq. (\ref{w1}) and (\ref{w2}) at $\beta=0$. The red and blue dots show the 
positions of fictitious A vortices which produce the current streamlines in the upper and the lower half-plane, respectively, as described in the text.}
\end{figure}

It turns out that the solution for AJ vortex is given by the complex potential $w_1(z)$ of a fictitious A vortex located at $z=u-il$, and $w_2(z)$ of another A vortex at $z=u+il$, where $u(t)$ is the position of the center of the AJ vortex core along the junction, as shown in Fig. \ref{ajcur}. This representation proposed for static and moving AJ vortices in an infinite junction \cite{AG,AG3} also works for AJ vortex in an overdamped  junction of finite length. To show this, we use a dimensionless complex potential of A vortex in a strip located at $0<x<1$: 
\begin{gather}
w_1(z)=i\ln\frac{\sin\frac{\pi}{2}(z-u+il)}{\sin\frac{\pi}{2}(z+u+il)}+\frac{i\beta x}{2\pi\epsilon}-\frac{\chi}{2},
\label{w1} \\
w_2(z)=i\ln\frac{\sin\frac{\pi}{2}(z-u-il)}{\sin\frac{\pi}{2}(z+u-il)}+\frac{i\beta x}{2\pi\epsilon}+\frac{\chi}{2},
\label{w2} 
\end{gather}
where $\chi(t)$ is a global phase difference between the superconductors on different sides of the junction. Using Eqs. (\ref{w1}) and (\ref{w2}), we obtain the local phase difference on the junction $\theta(x,t)=\varphi_2(x,-0)-\varphi_1(x,+0)$, and the current density in the film at $y>0$:
\begin{gather}
\theta=\chi+2\tan^{-1}\left[\frac{\sin\pi u\sinh\pi l}{\cos\pi u\cosh\pi l -\cos\pi x}\right],
\label{thet} \\
j_x-ij_y=\frac{2\pi^2i\epsilon \sin\pi u}{\cos\pi(z+il)-\cos\pi u}-i\beta,
\label{jj}
\end{gather}
where $\tan^{-1}(z)$ at $z<0$ is defined as $\pi-\tan^{-1}(|z|)$. 

Substituting Eqs. (\ref{thet})-(\ref{jj}) into Eq. (\ref{bc}), one can show that they are exact solutions for a moving AJ vortex in which $\chi(t)$, $u(t)$ and $l(t)$ satisfy the 
following ordinary differential equations (see Appendix B):
\begin{gather}
\tau\partial_t\chi+\sin\chi=\beta(t),
\label{e1} \\
i\tau\partial_t(u+il)=\frac{\sin\pi u\sinh\pi l }{\pi\sin\pi(u+il)}\exp(i\chi)-\pi\epsilon.
\label{e2} 
\end{gather}
Real and imaginary parts in Eq. (\ref{e2}) yield coupled cumbersome ODEs for $u(t)$ and $l(t)$ given in Appendix B. For a vortex being far away from the edges of a 
long junction, $l\ll 1$ and $u\sim 1$, Eq. (\ref{e2}) in normal units reduces to \cite{agac} 
\begin{gather}
\tau\partial_t u=l\sin\chi(t),
\label{a1} \\
\tau\partial_t l=-l\cos\chi(t)+l_0.
\label{a2}
\end{gather} 
For a dc current, $\dot{\beta}=\dot{l}=0$, Eqs. (\ref{a1}) and (\ref{a2}) yield Eq. (\ref{lv}) for the vortex velocity $v(\beta)$ and the core length $l(\beta)$.

Nonlinear Eqs. (\ref{e1})-(\ref{e2}) fully determine dynamics of the vortex position and the core length under the action of  
an arbitrary ac current $\beta(t)$. Here the coupled equations for $u(t)$ and $l(t)$ describe how the length of the core changes as  
it moves along the junction. This nonlinear effect is due to the change in the distribution of circulation currents and acceleration of the vortex as it approaches the edge of the junction. 
Here the equation for $\chi(t)$ turns out 
to be decoupled from $l(t)$ and $u(t)$, as it also occurs for the AJ vortex in an infinite junction \cite{agac}.     

\begin{figure} 
\includegraphics[width=\columnwidth]{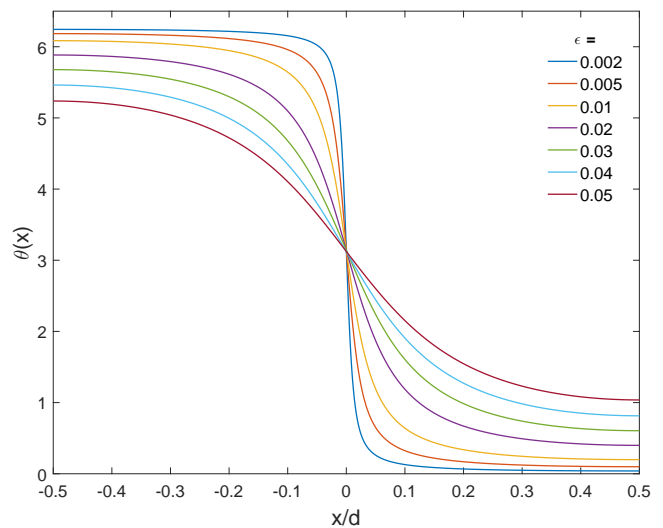}
\caption{\label{statsize} $\theta(x)$ in a static vortex calculated from Eq. (\ref{main}) for different values of $\epsilon$ as described in the text.}
\end{figure}

We also calculated $\theta(x)$ in a static AJ vortex by solving Eq. (\ref{main}) numerically at $\beta=0$ and the initial distribution of $\theta(x,0)=4\tan^{-1}\exp(-x/\epsilon)$ centered in the middle of the junction. To stabilize the vortex against attraction to the edges, a weak ``pinning" potential modeled by $J_c(x)=[1-\delta \exp(-x^2/\zeta^2)]J_c$ was incorporated.  Simulations of Eq. (\ref{main}) in which $\sin\theta$ is replaced with $[1-\delta \exp (-x^2/\zeta^2)]\sin\theta$ and $\delta=\zeta=0.02$ show that $\theta(x,0)$ evolves into stationary $\theta(x)$ presented in Fig. \ref{statsize} for different values of $\epsilon$. The so-calculated $\theta(x)$ coincides with $\theta(x)$ given by Eq. (\ref{thet}) with $\chi=0$ and $x\to x-1/2$ to the accuracy of the line width in Fig. \ref{statsize}, where 
\begin{equation}
\theta(x)=2\cos^{-1}\frac{\sin\pi x}{\sqrt{\sin^2\pi x+\sinh^2\pi l}}.
\label{tetstat}
\end{equation}
Here $\sinh^2\pi l =\pi^2\epsilon/(1-\pi^2\epsilon)$ is obtained from Eq. (\ref{st2}) at $\beta=0$. As follows from Fig. \ref{statsize} and Eq. (\ref{tetstat}), the vortex expands as 
$d$ decreases and $\epsilon$ increases. 

Shown in Figs. \ref{ajcur} and \ref{statsize} are current streamlines and $\theta(x)$ in the AJ vortex. 
Unlike vortices in a long junction which are $2\pi$ phase kinks with $\Delta\theta = \theta(-\infty)-\theta(\infty)=2\pi$, the AJ vortices in a short junction are partial phase kinks 
with $\Delta\theta<2\pi$. The latter reflects the fact that the AJ vortex carries a reduced magnetic flux $\phi<\phi_0$, as it is characteristic of vortices in thin films \cite{stej}. 
A phase shift produced by a real A vortex on a junction in a thin film strip was observed in Ref. \onlinecite{kras} 
and calculated in Ref. \onlinecite{KM}.
 
\subsection{Transition of AJ vortex into a phase slip}

Consider stationary solutions of  Eqs. (\ref{e1})-(\ref{e2}) for a dc current $\beta < 1$, that is $J<J_c$.
Setting the time derivatives to zero and separating real and imaginary parts of the right hand side of Eq. (\ref{e2}) yields $\sin\chi=\beta$, and:
\begin{gather}
\tan\left(\frac{\pi u}{d}\right)=\frac{\pi^2\epsilon}{\beta},
\label{st1} \\
\tanh\left(\frac{\pi l}{d} \right)=\frac{\pi^2\epsilon}{\sqrt{1-\beta^2}}.
\label{st2}
\end{gather}        
Equation (\ref{st1}) determines a stationary position of the vortex balanced by the Lorentz force of transport current and attraction of the vortex to the edges of the junction, which can be interpreted in terms 
of interaction of the vortex with a chain of V-AV images ensuring the boundary conditions $J_x(\pm d/2,0)=0$ at the edges.  
The position of AJ vortex given by Eq. (\ref{st1}) is unstable as a small displacement $\delta u(t)$ causes the vortex to move toward one of the film edges in a way similar to a stationary A vortex in a film \cite{stej}.  This also follows from the linear stability analysis given in Appendix B, which shows that small perturbations $\delta u(t)=\delta u(0)\exp(\gamma_u t)$ grow exponentially with the increment $\gamma_u=\pi^4\epsilon^2$ at $\beta=0$.     

Equation (\ref{st2}) which defines the length of AJ vortex core at the stationary position yields $l=l_0$ at $d\gg l_0$ \cite{AG}. However, for a junction of finite length, Eq. (\ref{st2}) has solutions only if $\pi^2\epsilon < \sqrt{1-\beta^2}$. Using here Eq. (\ref{eps}), we conclude that the stationary vortex solution exists only in a sufficiently long junction:
\begin{equation}
d>d_c=\frac{\pi l_0}{\sqrt{1-(J/J_c)^2}}.
\label{dc}
\end{equation}  
As $d$ approached $d_c$ from above, the AJ core length $l(J)$ in Eq. (\ref{st2}) diverges, and the stationary vortex solution (\ref{thet}) turns into a phase slip 
in which $\theta(x)$ is uniform along the junction. This result is in agreement with the numerical simulations shown in Fig. \ref{statsize} where the vortex spreads over the entire junction 
as $\epsilon$ approaches the critical value $\epsilon_c=\pi^{-2}$. The transition of a static AJ vortex into a phase slip at $d< d_c=\pi l_0$ resembles the ``core explosion" of 
a parallel A vortex in a film of thickness $d<d_c\simeq 3.6\xi$  which was obtained  by  numerical simulations of GL equations \cite{KL,vod}. For a  
perpendicular junction in a thin film shown in Fig. \ref{junc} (a), the condition $\pi l < d < \lambda$ that AJ vortex can exist while the London screening is negligible is satisfied if  
$l\ll \lambda$, that is $J_d/\kappa < J_c < J_d$. However, for an edge junction in a thin film, this condition $\pi l < d < 2\lambda^2/s$ becomes much less restrictive and can be satisfied in low-$J_c$ junctions. Notice that $d_c$ defined by Eq. (\ref{dc}) increases as the bias current increases.  

AJ vortex driven by any ac current in an overdamped junction does not radiate.  At $\eta\lesssim 1$ bremsstrahlung produced by the vortex due to its acceleration at the junction edges, and the Cherenkov radiation caused by NJE effects can give rise to a splitting instability of the vortex \cite{screp}. Results of numerical simulations of these effects which occur at $\eta\lesssim 1$ are presented in the next sections. 

\section{Dc current}
\label{sec:dc}

In this section we show results of simulation of Eq. (\ref{main}) for vortices driven by a dc current. We consider three situations: 1. Vortices penetrate from the edge of the junction where $J(x)$ exceeds $J_c$ due to a small gradient in $J(x)$ along the junction. 2.  Vortices appear inside the junction in a region where $J_c(x)$ is locally reduced. 3. Vortices appear due to coexistence of current gradient and a defect in the junction. Most of the simulations were done for $\epsilon=l_0/\pi d =2\cdot 10^{-3}$, that is, for long junctions much larger than the static AJ core size $l_0$.   

\subsection{Junction with weak screening} 
\label{sub:scrn} 
\begin{figure} 
\includegraphics[width=\columnwidth]{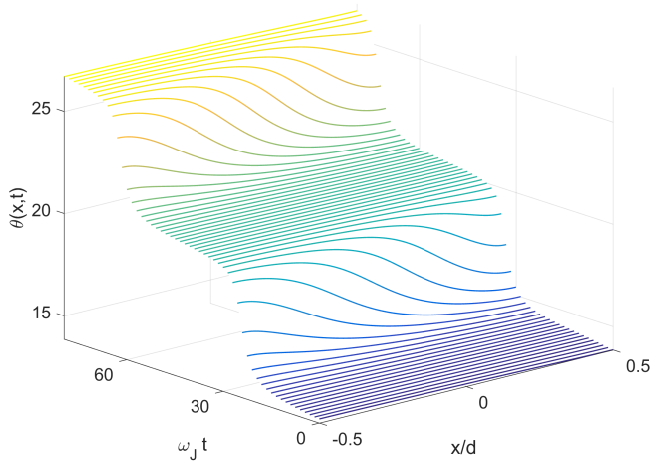}
\caption{\label{over3d} Penetration of single vortices in an overdamped junction with $\eta=2$ and $\beta_0=1.05$ calculated for $k=0.02$ and $\epsilon=2\cdot 10^{-3}$.}
\end{figure}

\begin{figure} 
\includegraphics[width=\columnwidth]{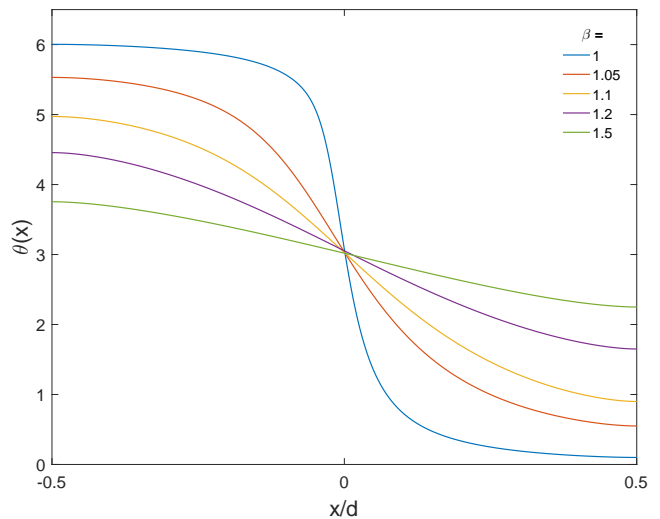}
\caption{\label{expprof} Snapshots of moving vortices in the middle of the junction calculated from Eq. (\ref{main}) for different currents at $\eta=2$, $k=0.02$ and $\epsilon=2\cdot 10^{-3}$.}
\end{figure}

Consider penetration of vortices in a junction, assuming that $\beta(x)=(1-kx)\beta_0$ in Eq. (\ref{main}) has a small gradient with $k\ll1$. The slight inhomogeneity in $J(x)$ with $k=d/\Lambda$ can result from self-field effects of transport current or a dc field applied to one side of a thin film screen with a perpendicular Josephson junction shown in Fig. \ref{junc}.  It turns out that the dynamic behavior of vortices in overdamped $(\eta \gtrsim1)$ and undedramped $(\eta \lesssim 1)$ junctions is markedly different. For $\eta \gtrsim 1$, simulations of Eq. (\ref{main}) with $\beta(x)=(1-kx)\beta_0$ show that, once $J$ exceeds $J_c/(1+k)$, vortices start penetrating one by one through the left edge of the junction and exiting from the other end (Fig. \ref{over3d}). Figure \ref{expprof} shows that as $J$ increases, the flight time of vortices through the junction decreases while the size of a vortex increases. The expansion of moving J and AJ vortices as $\beta_0$ increases is characteristic of the overdamped limit \cite{KL,AG} (see also Eqs. (\ref{ajkink}) and (\ref{lv})).  Based on the results presented above, we can therefore expect a transition of moving vortices into a phase slip as the current increases even in a long junction with $d>d_c$ where a static vortex can exist.  

Our numerical simulations of Eq. (\ref{main}) with $\eta\gtrsim 1$ have shown that a gradual transition of a moving vortex into a phase slip does happen as $\beta_0$ increases and   
the vortex spreads over the entire junction. In this case $\theta(x,t)$ becomes flat and increases nearly linearly with $t$. 
For $\beta_0\gg 1$, the phase slip state $\theta(x,t)$ is described by 
\begin{equation}
\theta(x,t)=\theta_0(t)+\delta\theta(x,t),
\label{pst}
\end{equation}
where $\theta_0(t)$ satisfies the equation for a point contact:
\begin{equation}
\ddot{\theta}_0+\eta\dot{\theta}_0+\sin\theta_0=\beta_0.
\label{t0t}
\end{equation} 
For $\beta_0\gg 1$ and $\eta\gg 1$, an approximate solution of Eq.(\ref{t0t}) is:
\begin{equation}
\theta_0(t)=\frac{\beta_0t}{\eta}+\frac{\eta^2}{\beta_0^2+\eta^4}\left[\sin\frac{\beta_0t}{\eta}+\frac{\eta^2}{\beta_0}\cos\frac{\beta_0t}{\eta} \right]
\label{t0}
\end{equation}

\begin{figure} 
\includegraphics[width=\columnwidth]{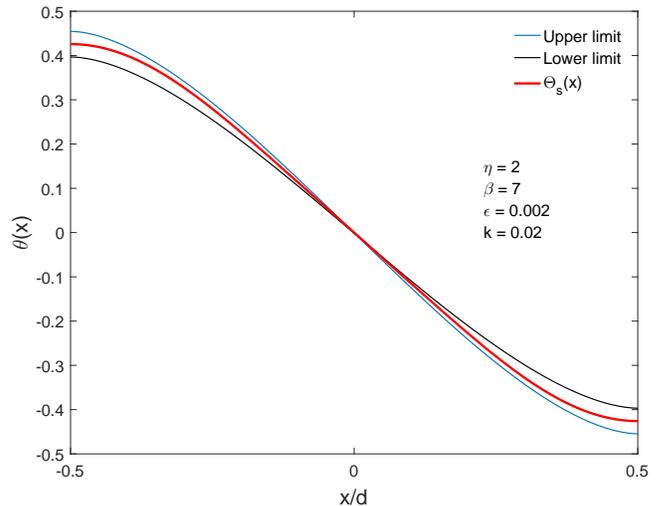}
\caption{\label{unipro} Upper and lower limits between which $\theta(x,t)$ oscillates, calculated for $\eta=2$ and $\beta_0=7$. The red curve shows $\theta_s(x)$ described by Eq. (\ref{uni}).}
\end{figure}

A small correction $\delta\theta(x,t)$ in Eq. (\ref{pst}) comes from the integral and the nonlinear terms in Eq. (\ref{main}).  Figure \ref{unipro} shows that  
the calculated $\delta\theta(x,t)$ oscillates around a stationary profile $\theta_s(x)$ caused by the weak inhomogeneity of $\beta(x)=(1-kx)\beta_0$ (see Appendix A):
\begin{equation}
\theta_s(x)=-\frac{4k\beta_0}{\pi^4\epsilon}\sum_{n=o}^{\infty}\frac{(-1)^n\sin\pi(2n+1)x}{(2n+1)^3}
\label{uni}
\end{equation}
 
\begin{figure} 
\includegraphics[width=\columnwidth]{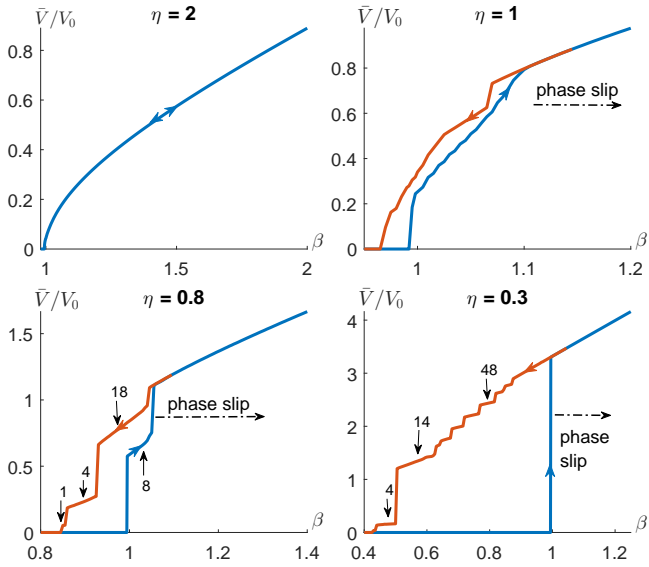}
\caption{\label{dciv} The dc voltage $\bar{V}=\langle V(t)\rangle$ calculated from Eq. (\ref{fur}) for different values of $\eta$, where $\langle ... \rangle$ denotes time averaging, 
and $V_0=\phi_0\omega_J/2\pi c$.}
\end{figure}

To see how the gradual transition from the vortex to the phase slip state can manifest itself in the $V-I$ characteristics, we calculated the averaged instantaneous voltage on the junction:
\begin{equation}
V(t)=\frac{\phi_0\omega_J}{2\pi c}\int_{-1/2}^{1/2}\dot{\theta}(x,t)dx=\sum_\omega V_\omega\exp(i\omega t).
\label{fur}
\end{equation}
Here $V(t)$ has multiple Fourier harmonics caused by superposition of Josephson oscillations and motion of vortices.  The behavior of AJ vortices in a long junction can be inferred from the dc component of voltage $\bar{V}(\beta_0)$ shown in Fig. \ref{dciv}.  At $\eta=2$ the calculated $V-I$ curve follows $V=I_cR\sqrt{\beta_0^2-1}$ for the overdamped point junction \cite{BP} for all $\beta_0$ except for a vicinity of $\beta_0\approx 1$ where the phase slip transition occurs.  At $\eta\lesssim 1$ the $V-I$ curves acquire stepwise features and become hysteretic. Here the jumps in the ascending branches of $\bar{V}(\beta_0)$ result from penetration of several vortices which then turn into a phase slip state at larger $\beta_0$ indicated by the dashed arrows.  The descending branches of $\bar{V}(\beta_0)$ exhibit staircase structures where steps correspond to different numbers of vortices indicated by vertical arrows.


\begin{figure} 
\includegraphics[width=\columnwidth]{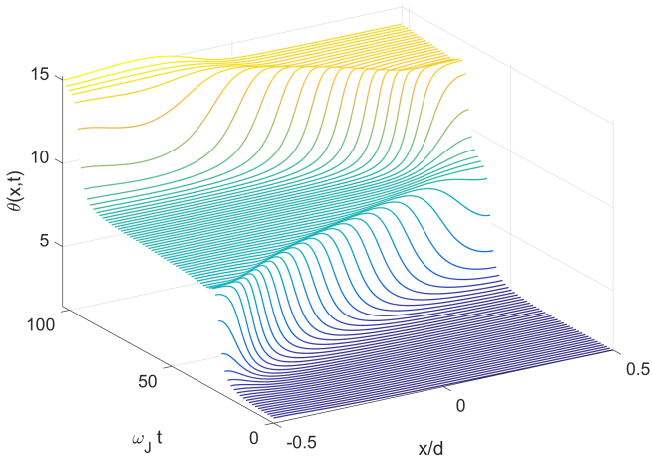}
\caption{\label{med3d} A wake radiated behind the moving vortex at $\eta=1$ and $\beta_0=0.995$. Here the vortex gets reflected from the edge and turns into antivortex. }
\end{figure}

\begin{figure} 
\includegraphics[width=\columnwidth]{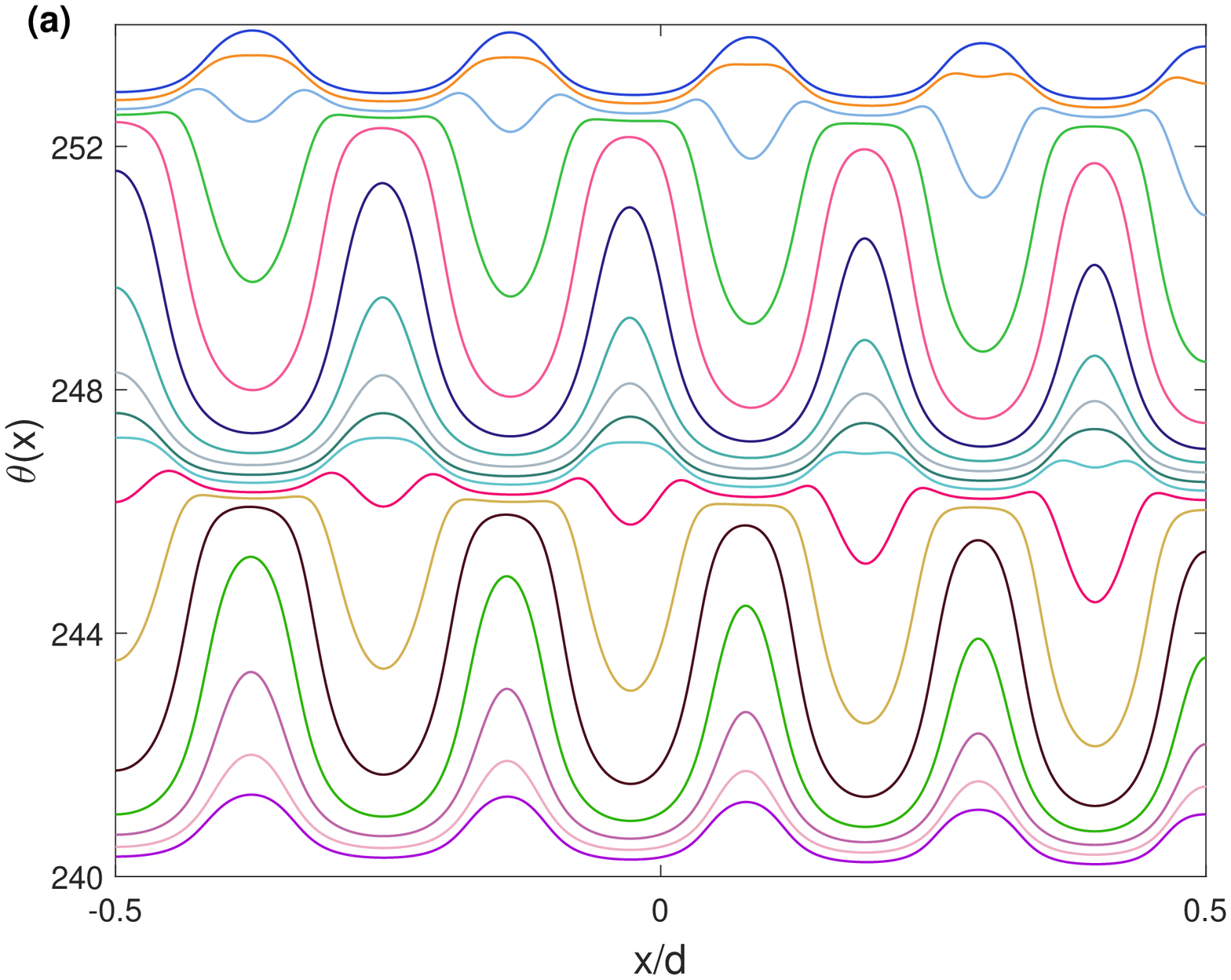}
\includegraphics[width=\columnwidth]{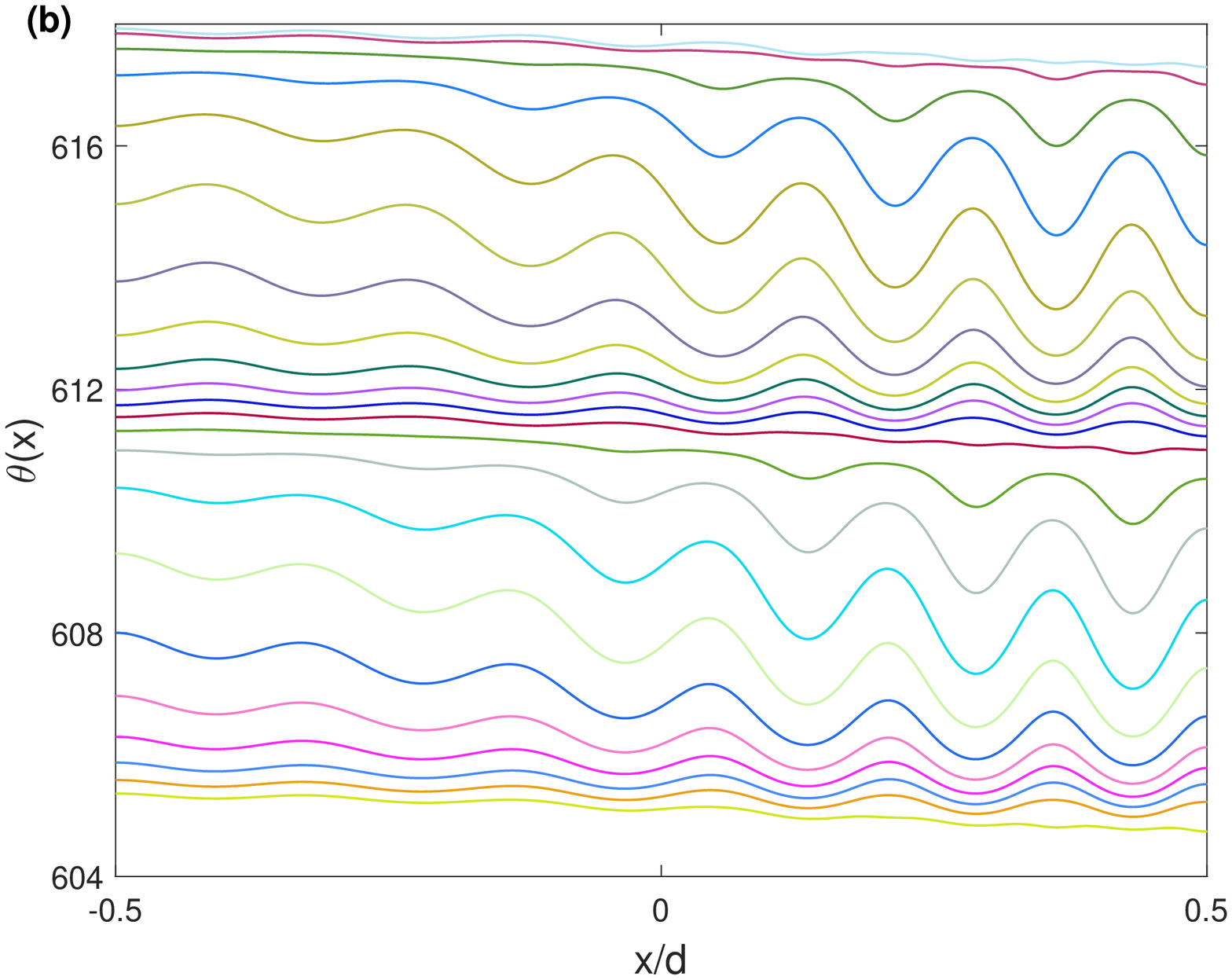}
\caption{\label{stnd} Snapshots of dynamic patterns formed by counter-moving vortices and antivortices calculated for $\eta=1$, $\epsilon = 2\cdot 10^{-3}$, $k=0.02$, $\beta_0 = 1.05$ (a) and $\beta_0 = 1.09$ (b). Different colors correspond to different times $t$ during the time period after which the phase structures repeat themselves periodically after shifting up in $\theta$. As current further increases, the patterns shown in Figs. \ref{stnd} (a) and (b) gradually turn into a phase slip profile similar to that is shown in Fig. \ref{unipro}. The asymmetry of $\theta(x,t)$ with respect to $x=0$ is due to the effect of the gradient in $\beta(x)$.}
\end{figure}

\begin{figure} 
\includegraphics[width=\columnwidth]{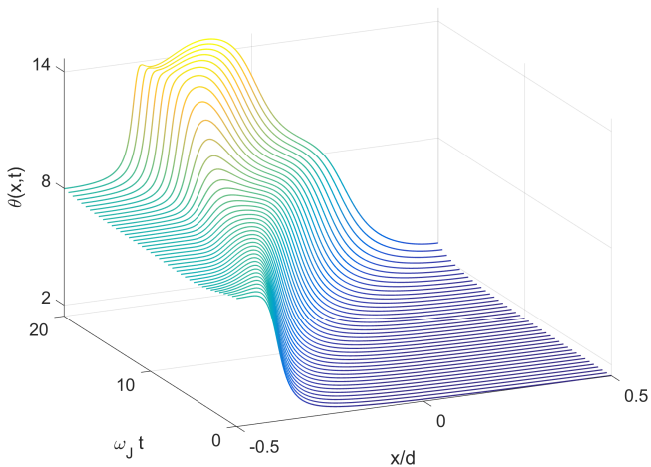}
\caption{\label{split3d} Initial state of generation of V-AV pairs calculated at $\eta=0.7$ and $\beta_0=0.995$. The Cherenkov splitting instability of a vortex occurs right after it enters the junction 
and ultimately results in the dynamic pattern similar to those shown in Fig. \ref{stnd}.}
\end{figure}

The behavior of $\bar{V}(\beta_0)$ on the ascending branch is illustrated by  Figs. \ref{over3d} -\ref{expprof} and  \ref{med3d}-\ref{split3d} which show representative $\theta(x,t)$ calculated for different values of $\eta$ and $\beta_0$. In an overdamped junction $(\eta\gtrsim 2)$ vortices periodically appear at the left edge, move along the junction and disappear at the right edge. As $\beta_0$ increases vortices move faster and become longer, which eventually results in the transition to the phase slip state described above (see Fig. \ref{over3d}). In this case strong ohmic dissipation suppresses both the Cherenkov radiation caused by the nonlocal effects and bremsstrahlung resulting from acceleration and deceleration of a vortex as it moves along the junction. This behavior of vortices starts changing at $\eta\approx1$ as the radiation wake behind a moving vortex shown in Fig.  \ref{med3d} becomes apparent. In this case vortices which reach the edge of the junction get reflected as vortices of opposite polarity (antivortices). As a result, vortices penetrating from the left edge of the junction collide with antivortices reflected from its right edge: at $\eta\lesssim 1$ these vortices and antivortices do not annihilate but go through each other, similar to underdamped Josephson vortices described by the sine-Gordon equation \cite{BP}. As current further increases, the number of vortices and antivortices in the junction increases and eventually counter-moving vortices and antivortices form a dynamic pattern shown in Fig. \ref{stnd}. This state can be regarded as a nonlinear wave on the background phase $\theta_0(t)$ which increases with time, so that the snapshots of $\theta(x,t)$ shown in Fig. \ref{stnd} shift up and periodically replicate themselves. As the current increases, the overlap of vortices and antivortices reduces the amplitudes of the phase waves as shown in Fig. \ref{stnd} (b). As the current increases further, this structure which manifests itself in the behavior of $\bar{V}(\beta_0)$ at $1<\beta_0<1.15$, turns into a phase slip state, shown in Fig. \ref{dciv}. 
 
At $\eta=0.9$ the first signs of vortex splitting instability caused by the Cherenkov wake behind the vortex penetrating from the left edge appear. As the vortex approaches the right edge it accelerates due to attraction to the edge so that the wake amplitude increases and exceeds a critical value above which a V-AV pair forms. The junction eventually goes into a dynamic steady-state after two more V-AV pairs are generated at the edges. This Cherenkov instability becomes more apparent at $\eta=0.8$ for which the wake amplitude exceeds the threshold when the vortex reaches the middle of the junction where a V-AV pair first appears. The newborn vortex and antivortex move apart, accelerate and produce another V-AV pair. These vortices with opposite polarities oscillate back and forth in the junction and form a dynamic structure similar to that is shown in Fig. \ref{stnd}.  At a slightly higher current more V-AV pairs are generated and the junction goes into the phase slip state. 

At $0.3<\eta<1$ dynamic multi-vortex structures on the ascending branch of $\bar{V}(\beta_0)$ exist in a narrow range of currents $(1+k)^{-1}<\beta_0<\beta_s$ which shrinks as $\eta$ decreases and vanishes at $\eta = 0.3$ at which the phase slip current $\beta_s=(1+k)^{-1}$. Vortices at $\eta <0.3$ exist only during a transient period during which the junction goes into a phase slip state after the current density at the edge reaches the threshold of vortex penetration. For instance, our simulations of Eq. (\ref{main}) at $\eta=0.2$ showed that, once a vortex enters the junction, it produces a V-AV pair which in turn triggers a cascade of V-AV pairs driving the junction into a resistive phase slip state. This behavior is similar to the phase pile expansion \cite{screp} shown in Fig. \ref{scireppro}. Simulation videos of the dynamics of the junctions in different regimes are available at Ref.\onlinecite{supp}. The Cherenkov instability of vortex right after it enters through the edge of the junction and the subsequent transition to a resistive state manifests itself in big jumps on the ascending branches of $\bar{V}(\beta_0)$ shown in Fig. \ref{dciv} for $\eta=0.3$. However, the subsequent decrease of current results in re-appearance of vortices from the phase slip state, which manifests itself in the hysteresis in the $V-I$ curves and the staircase form of the descending branch of $\bar{V}(\beta_0)$. The evolution of non-hysteretic $V-I$ curves to hysteretic ones upon decreasing $\eta$ in a long junction considered here resembles the well-known transition from non-hysteretic to hysteretic $V-I$ curves in point junctions \cite{BP,KL}, except that the returned descending branch of $\bar{V}(\beta_0)$ in Fig. \ref{dciv} is controlled by vortices emerging from the phase slip state.       

\begin{figure} 
\includegraphics[width=\columnwidth]{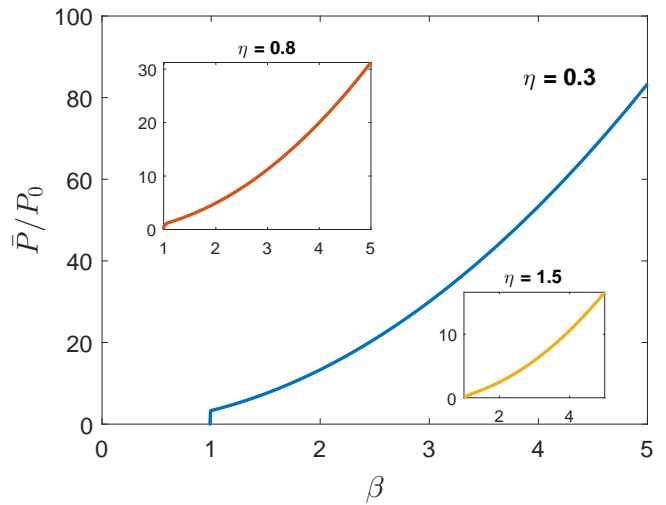}
\caption{\label{dcdiss} Dissipation power vs dc current calculated for different damping constants shows a quadratic behavior at currents well above the threshold of penetration of a vortex.}
\end{figure}

Shown in Fig. \ref{dcdiss} is the power $\bar{P}(\beta_0 )$ dissipated by moving vortices calculated from Eqs. (\ref{main}) and (\ref{diss}) for different $\eta$. The curves $\bar{P}(\beta_0)$ have jumps and hystertic features at the onset of vortex penetration which reflect those in Fig. \ref{dciv}. However, once $\beta_0$ exceeds the phase slip transition threshold, the dependence of $\bar{P}$ on $\beta_0$ nearly follows that of a point junction and exhibits the ohmic quadratic behavior $\bar{P}=\beta_0^2P_0/\eta$ at large $\beta_0$. The latter is similar to $\bar{P}(\beta_0)$ for Josephson vortices in a long junction described by the sine-Gordon equation \cite{physc}. 

\begin{figure} 
\includegraphics[width=\columnwidth]{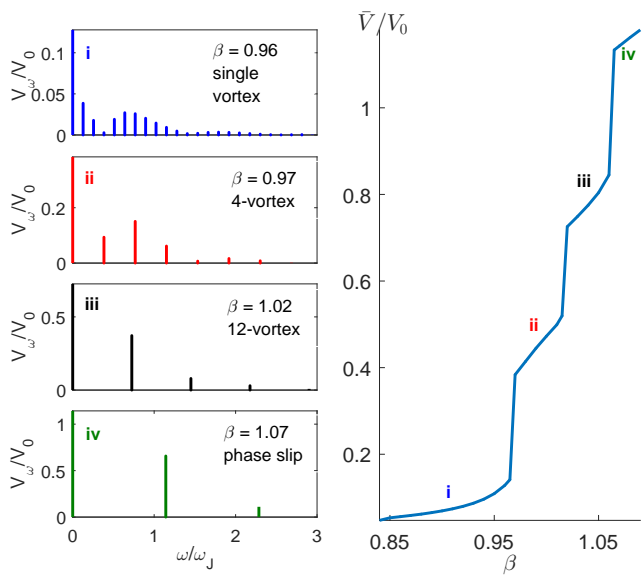}
\caption{\label{fspec} Spectrum of Fourier components of voltage $V(t)$ calculated for $\eta=0.8$ and different currents corresponding to different number of vortices in the junction (left panel). Right panel shows the dc voltage $\bar{V}(\beta)$ in which jumps result from the change of the number of vortices in the junction.}
\end{figure}

Transitions between different dynamic vortex patterns can also manifest themselves in the voltage Fourier spectrum in Eq. (\ref{fur}). We calculated the Fourier spectrum   
by solving Eq. (\ref{main}) with a uniform current $\beta$ and $\eta=0.8$, using the static solution (\ref{tetstat}) as the initial condition. It turned out that if $\beta < 0.84$, the vortex 
is pushed by the current to the edge of the junction and exits. However at $\beta > 0.85$, the vortex gets trapped in the junction as it starts bouncing back and forth between the edges and interacting with radiated waves it produces. Then the current was incrementally increased to $\beta+\Delta\beta$ and Eq. (\ref{main}) was solved using the calculated solution at the preceding $\beta$ as the initial condition. Above a threshold current this single vortex produces a V-AV pair, forming a periodically changing structures of vortices and antivortices glued by Cherenkov radiation, similar to those shown in Fig.  \ref{stnd}. In this way the dc voltage $\bar{V}(\beta)$ shown in the right panel of  Fig. \ref{fspec} was obtained. The so-calculated $\bar{V}(\beta)$ has jumps corresponding to the current-driven transitions between different number of vortices in the junction. Using the solution $\theta(x,t)$ we calculated the amplitudes $V_\omega$ of the Fourier harmonics
$$
V_\omega=\frac{V_0}{T}\left|\int_0^T dt e^{-i\omega t}\int_{-1/2}^{1/2}\dot{\theta}(x,t)dx\right|,
$$  
where $T$ is the period of oscillations, and $V_0=\phi_0\omega_J/2\pi c$. The left panel in Fig. \ref{fspec} shows the voltage Fourier spectra at different $\beta$ corresponding to different number of 
vortices in the junction.  As the current increases and junction goes from a multi-vortex to the phase slip state, the amplitudes of low-frequency Fourier components $V_\omega$ 
with $\omega < \omega_J $ gradually diminish and finally disappear.  

\subsection{Penetration of vortices at the edge defect}
\label{sub:point}

\begin{figure} 
\includegraphics[width=\columnwidth]{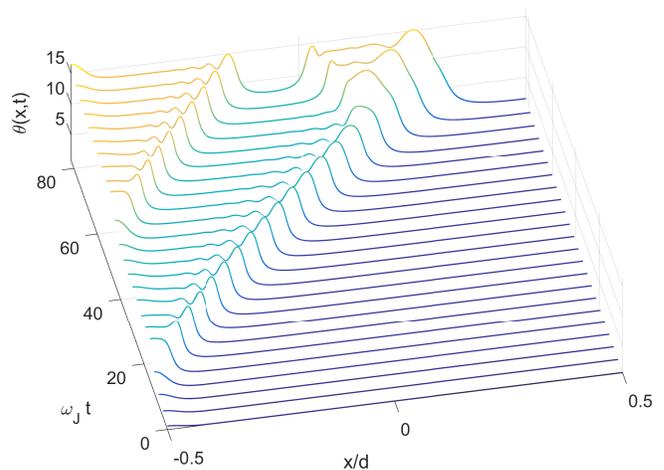}
\caption{\label{po3d} A vortex depinned from the defect at the left edge of the junction accelerates and produces a V-AV pair at $x\approx 0.1$ after the next vortex enters the junction. Simulations were done for $\delta_0=0.5$, $\beta=0.8$ and $\eta=0.3$.}
\end{figure}

Penetration of vortices in the junction can be facilitated not only by a weak gradient in $\beta(x)$, but also by a small defect at one of the edges. Such defects which are common in thin film junctions can locally reduce the Josephson critical current density $J_c(x)$. This situation can be modeled by Eq. (\ref{main})  in which   
\begin{gather}
\sin\theta\to [1-f(x)]\sin\theta, 
\label{sinh} \\
 f(x)=\delta_0\exp\left[-\frac{(x+1/2)^2}{\zeta^2}\right].
 \label{delt}
\end{gather}
Here $\delta_0=\delta J_c(-d/2)/J_c$ quantifies the magnitude of the local reduction of $J_c(x)$ at the edge, and  $\zeta$ is a dimensionless length of the defect.  
In our simulations we set $\zeta = 0.05$ and assumed that $\beta$ is uniform. The results show that at $\eta>1$ vortices penetrate one by one, their size expands as current increases and the transition to the phase slip state occurs. At $\eta<1$ vortices get reflected from the edges and the radiation wake behind moving vortices becomes apparent.  Further increase of $\beta$ yields dynamic structures similar to those shown in Fig. \ref{stnd} and their subsequent transition to the phase slip state.  At $\eta<0.3$ a vortex depinned from the edge defect by current accelerates and produces enough radiation to generate a V-AV pair which then multiplies and drives the entire junction into the resistive phase slip state.  An example of such transient state is shown in Fig. \ref{po3d} in which the first V-AV pair appears as the initial vortex traveled more than half the length of the junction. Our detailed simulations of dynamics of vortices in the presence of edge defects have shown that the threshold current for vortex penetration decreases as the size of the defect increases \cite{supp}. The apparent similarity of the dynamics of vortices for the cases of edge defect and current gradient suggests that the transition to the phase slip state in both cases is mostly controlled by the values of $\eta$ and $\beta$.

\subsection{Interaction of vortices with pinning centers in the junction}
\label{sub:vp}
\begin{figure} 
\includegraphics[width=\columnwidth]{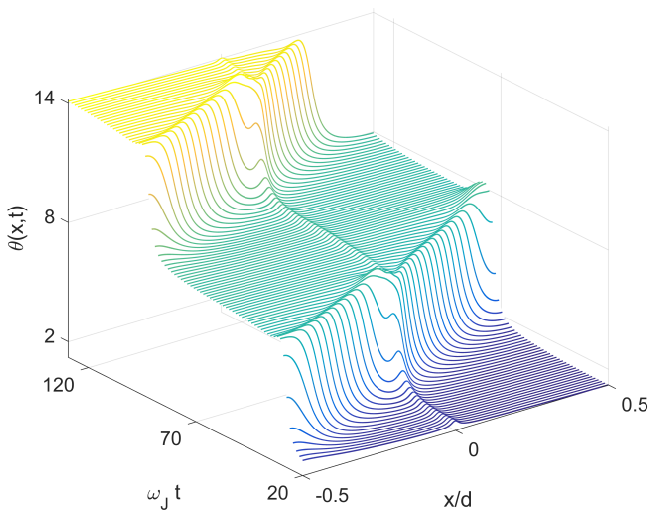}
\caption{\label{wkdef} A vortex accelerates as it approaches the defect in the center and decelerates once it passes the defect in the case of $\zeta=0.01$, $\delta_0=0.15$, $\beta_0=0.98$, and $\eta=1$.}
\end{figure}

\begin{figure} 
\includegraphics[width=\columnwidth]{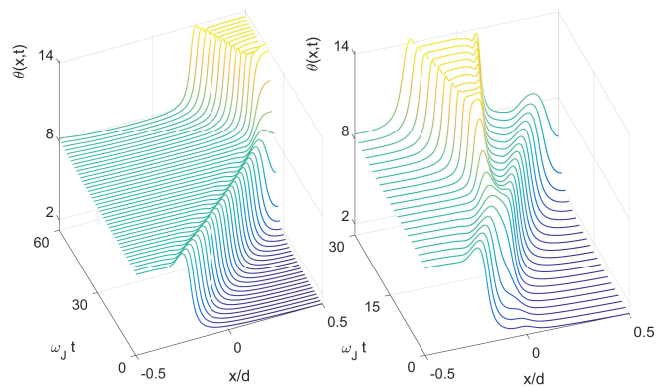}
\caption{\label{ddins} At $\eta=0.7$ even a weak defect can accelerate the approaching vortex so that it produces a critical radiation wake which generates a V-AV pair. Figure shows the dynamics of a vortex in the absence (left) and the presence (right) of a defect with $\delta_0=0.05$, $\zeta=0.05$ and $\beta_0=0.98$.}
\end{figure}

Consider now a moving vortex interacting with a defect in the middle of the junction in which case $f(x)$ in Eqs. (\ref{sinh}) and (\ref{delt}) is modeled by a Gaussian peak centered at $x=0$.  Let a vortex enter from the left edge of the junction due to a weak current gradient $\beta(x)=(1-kx)\beta_0$ with $k=0.1$, as was considered in subsection \ref{sub:scrn}. We focus here on strong currents $\beta_0\gtrsim 1$ for which the defect is too weak to pin the vortex, yet the dynamics of vortices can change substantially,  depending on the values of $\delta_0$ and $\zeta$. Shown in Fig.  \ref{wkdef} are the results of simulations for a weak defect with $\delta_0=0.15$ and $\zeta=0.01$ at $\eta=1$. Here the vortex enters from the left edge of the junction, accelerates and decelerates as it approaches and passes the defect, and then accelerates again as it exits from the right edge. Dynamics of the vortex can change markedly if $\eta$ is reduced and the radiation effects become essential. For instance, in the case of $\eta=0.7$ shown in Fig. \ref{ddins}, the Cherenkov wake increases as the vortex accelerates toward the defect, the wake amplitude exceeds the critical value at which the vortex produces a V-AV pair as it passes through the defect.  At smaller $\eta$ the vortex penetrating from the edge starts generating V-AV pairs before it reaches the defect, and the rest of dynamics is similar to what has been described in subsection \ref{sub:scrn}.

\begin{figure} 
\includegraphics[width=\columnwidth]{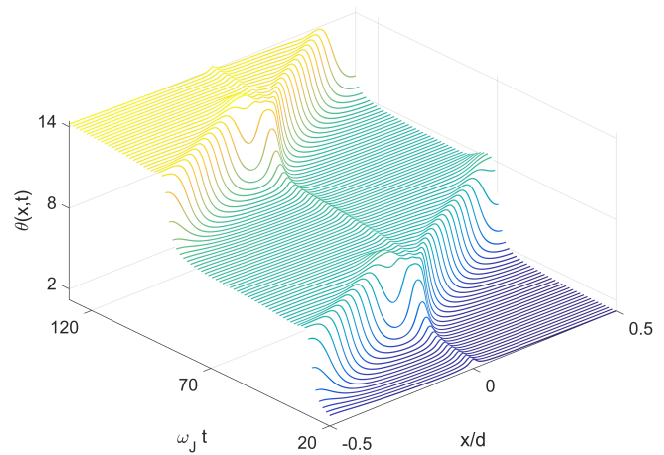}
\caption{\label{meddef} Interaction of a vortex penetrating from left with a V-AV pair produced simultaneously by a weak defect with $\delta_0=0.2$ and $\zeta=0.01$, $k=0.1$ and $\eta=1$ at the threshold current $\beta_0=0.98$. The vortex which entered from the left edge annihilates with the antivortex produced at the defect, and the remaining vortex exits from the right edge.}
\end{figure}

If $\delta_0=0.2$ and $k=0.1$, a vortex penetrates from left and simultaneously a V-AV pair appears at the defect. The subsequent dynamics of this vortex state depends on the values of $\eta$ and $\beta_0$. For instance, at $\eta=1$ and $\beta_0=0.98$, the vortex penetrating from left annihilates with the antivortex produced at the defect in the middle of the junction, while the remaining vortex exits from the right edge, as shown in Fig. \ref{meddef}. However, for the same parameters at larger current $\beta_0=0.995$, vortex and antivortex go through each other. Defects with $\delta_0 > 0.2$ and $\zeta = 0.01$ can trigger generation of V-AV pairs in the middle of the junction at a critical value $\beta_0\approx 1$ before any vortex enters from edges. In this case dynamics of vortices depends on $\eta$ in the same way as for the edge defect discussed in subsection \ref{sub:vp}. For a uniform current $(k=0)$, penetration of vortices at the edge defect can be mapped onto generation of V-AV pairs at the bulk defect in the region $0<x<0.5$, the two cases become equivalent if the length of the junction for the edge defect is reduced by half, that is, the parameter $\epsilon$ is doubled.  

\section{Ac current}
\label{sec:ac}

Consider now vortices driven by ac current with a small gradient in $\beta(x,t)=\beta_0(1-kx)\sin\omega t$, where $\omega$ is the dimensionless frequency in units of $\omega_J$. The results presented below were obtained for $\omega=\pi/30$.  Dynamics of vortices under ac current has several distinctive features as compared to the dc current: 
\begin{enumerate}
\item Since $\beta(t)$ changes sign periodically, penetration of vortices from the left edge is followed by penetration of antivortices. Vortices and antivortices produced during positive and negative cycles of $\beta(t)$ collide and either annihilate or produce bursts of radiation inside the junction. 
\item Vortices only penetrate during parts of the ac period when $\beta(t)=\beta_0\sin\omega t$ exceeds the penetration threshold $\beta_c$.  Our results show that $\beta_c$ depends on both $\omega$ and $\eta$: for instance, $\beta_c$ decreases from $1.22$ at $\eta=2$ to $1.01$ at $\eta=0.2$. 
\item Acceleration and deceleration of vortices under ac current bring about one more source of radiation which contributes to the generation of V-AV pairs.  
\item Dynamics of vortices under ac current changes markedly if the amplitude of oscillations of a vortex exceeds the length of the junction. 
\item Resonance interaction of oscillating vortices with standing waves in the junction affects the transition from vortices to phase slips and the generation of V-AV pairs. Analysis of these issues requires 
taking into account intertwined effects of  $\eta$, $\omega$, $\beta_0$, and $d$ on the dynamics of $\theta(x,t)$. 
\end{enumerate}
Given the complexity of ac dynamics of vortices affected by many different parameters, we only outline here a few essential cases (see Ref. \onlinecite{supp} for more details).   

\begin{figure} 
\includegraphics[width=\columnwidth]{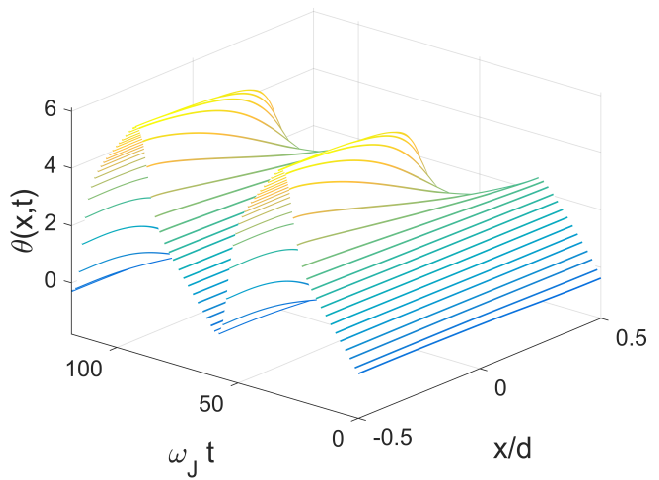}
\caption{\label{ac3d} Oscillatory dynamics of vortices in an overdamped junction with $\eta=2$ at the penetration threshold $\beta_0=1.237$. The vortex enters the junction during the positive cycle of $\beta(t)$, stops midway when $\beta(t)=0$, turns around and exits through the edge during the negative cycle of $\beta(t)$. }
\end{figure}

\begin{figure} 
\includegraphics[width=\columnwidth]{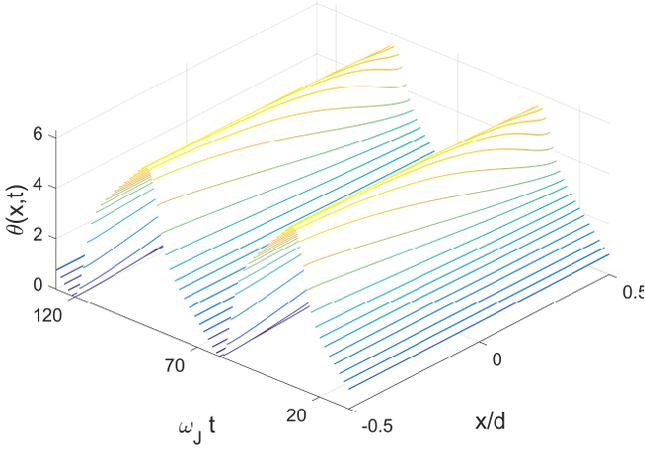}
\caption{\label{acex3d} Ballistic penetration of vortices and antivortices into an overdamped junction with $\eta=2$ at $\beta_0=1.245$.  Here vortices and antivortices traverse the junction and exit from the other end. Notice that the moving vortex extends nearly over the entire junction and produces no visible radiation.}
\end{figure}

\begin{figure} 
\includegraphics[width=\columnwidth]{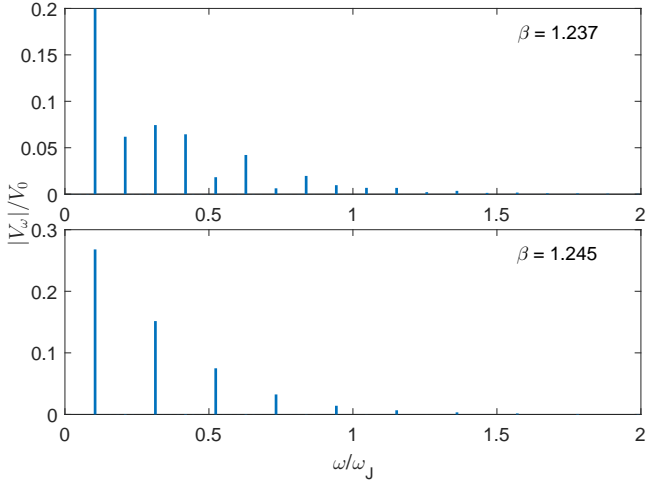}
\caption{\label{fcomp} Fourier spectrum of voltage at $\eta=2$ calculated for oscillatory vortex dynamics at $\beta_0=1.237$ and ballistic vortex penetration at $\beta_0=1.245$ represented in Figs. \ref{ac3d} and \ref{acex3d}, respectively. The peaks in $V_\omega$ occur at the multiples of the ac frequency $\omega_n=n\omega$, where $\omega=\pi/30$ and $n=1,2,3, ... $ . Notice that voltage harmonics with even $n$ disappear as the vortex dynamics changes from oscillatory to ballistic.}
\end{figure}

\begin{figure} 
\includegraphics[width=\columnwidth]{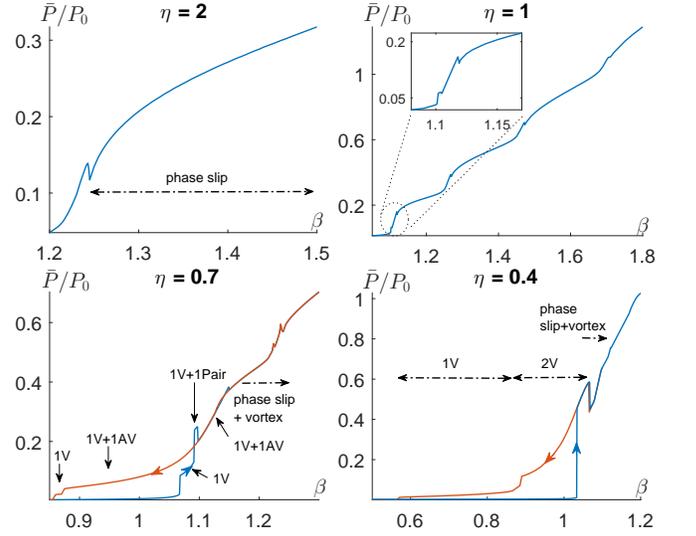}
\caption{\label{acprofiles} AC power plots $\bar{P}(\beta_0)$ for different damping constants. At large currents $\beta_0\gtrsim 3$, the curves $\bar{P}(\beta_0)$ approach the ohmic limit $\bar{P}=P_0\beta_0^2/2\eta$.}
\end{figure}

Figure \ref{ac3d} shows $\theta(x,t)$ calculated at $\eta=2$, $\beta_0=1.237$ and $\omega=\pi/30$. In this case a vortex enters the junction once $\beta(t)$ exceeds $\beta_c$ but, as $\beta(t)$ changes sign, the vortex turns around and exits through the same edge of the junction during the negative ac cycle, after which the whole process repeats. Neither antivortices nor radiation behind the moving vortex is visible here. However, at a slightly larger current $\beta_0=1.245$ the vortex expands further and becomes faster, so it  can move all the way to the other end of the junction and exit before $\beta(t)$ changes sign. During the negative ac cycle the antivortex enters the junction in the same way and extinguishes the positive phase shift left behind the preceding vortex, as shown in Fig. \ref{acex3d}.  The transition from the oscillating to the ballistic vortex dynamics manifests itself in the Fourier spectrum of voltage shown in Fig. \ref{fcomp}. In the oscillatory state the Fourier spectrum consists of equidistant peaks at $\omega_n=n\omega$, where $\omega=\pi/30$ and $n=1,2,3, ... $, while in the ballistic state the harmonics with even $n$ disappear. This transition also manifests itself in a negative jump in the dissipated power $\bar{P}(\beta_0)$ at $\beta_0\approx 1.245$, as shown in Fig. \ref{acprofiles}. Such $N$-shaped dependence of $\bar{P}(\beta_0)$ indicates a negative differential resistance and a hysteretic switching of the junction between two ascending branches of $\bar{P}(\beta_0)$ as the ac current amplitude is varied around $\beta_0\approx 1.245$. Here the phase slip state emerges at $\beta_0\geq 1.245$.   

\begin{figure} 
\includegraphics[width=\columnwidth]{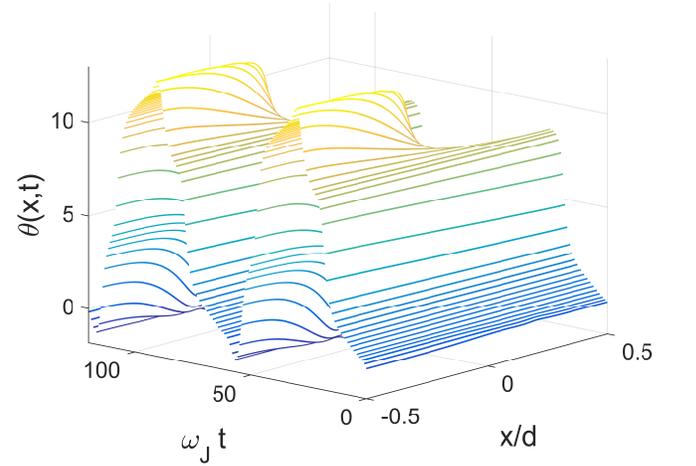}
\caption{\label{actovx} Partial penetration of a second vortex during positive cycle on top of ballistic penetration of first vortex which results in the second step in $\bar{P}(\beta_0)$ curve at $\eta=1$ in Fig. \ref{acprofiles} calculated for $\beta_0=1.26$.}
\end{figure}

Behavior of vortices becomes more complex as $\eta$ is decreased. For instance, at $\eta=1$, the curve $\bar{P}(\beta_0)$ shown in Fig. \ref{acprofiles} (b) acquires a staircase shape, each step resulting from penetration of an additional vortex. Close to the voltage onset at $\beta_0 = 1.102$ a vortex partially penetrates the junction during the positive cycle, then exits during the negative cycle, after which an anti-vortex partially enters and exits as the current changes sign again. This symmetry of the V-AV penetration breaks as current increases, so that a vortex penetrates deep into the junction during the positive ac cycle and returns during the negative ac cycle, but the antivortex does not penetrate, similar to the case shown in Fig.  \ref{ac3d} for $\eta=2$. As $\beta_0$ increases dynamics of a vortex changes from oscillating to ballistic, resulting in a $N-$shaped feature in $\bar{P}(\beta_0)$ at $\beta_0\approx 1.118$. At $\beta_0>1.118$, the ballistic penetration of vortices and antivortices proceeds in a way similar to that is shown in Fig. \ref{acex3d} until the appearance of the next step on the $\bar{P}(\beta_0)$ corresponding to the penetration of an additional vortex. In this case one vortex moves ballistically along the junction followed by a partial penetration of a second vortex. As current changes sign, this second vortex exits through the left edge followed by ballistic penetration of an antivortex, extinguishing the $4\pi$ phase shift acquired during the positive ac cycle (Fig. \ref{actovx}). As $\beta_0$ increases further, the transition from the oscillatory to ballistic dynamics of the second vortex also manifests itself in a small N-shaped feature in $\bar{P}(\beta_0)$ at $\beta_0\approx 1.269$ in Fig. \ref{acprofiles} (b). It turns out that, except for the small $N-$shaped features due to the transitions from oscillatory to ballistic dynamics of vortices, the curve $\bar{P}(\beta_0)$ calculated from Eq. (\ref{main}) for $\eta=1$ is close to $\bar{P}(\beta_0)$  of a point Josephson junction.

\begin{figure} 
\includegraphics[width=\columnwidth]{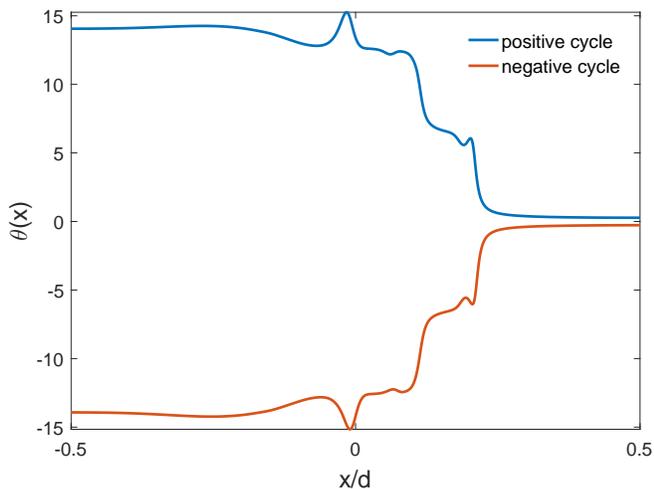}
\caption{\label{acrad} Penetration of radiating vortices and antivortices at $\eta=0.4$ and $\beta_0=1.034$.}
\end{figure}

\begin{figure} 
\includegraphics[width=\columnwidth]{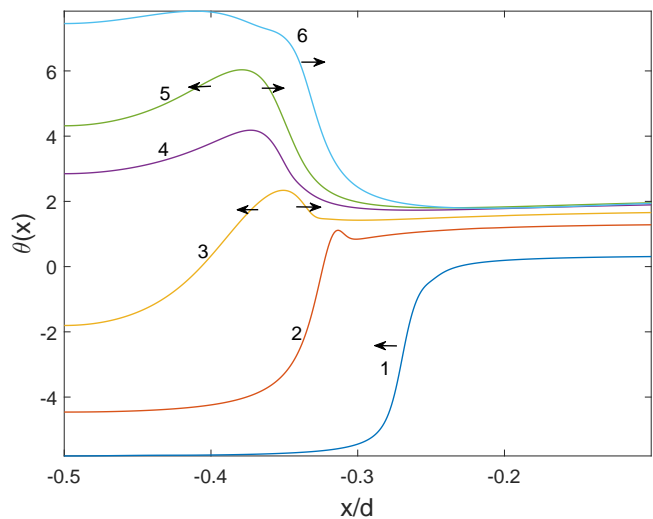}
\caption{\label{acvav} Generation of V-AV pair by the accelerating antivortex exiting the junction at $\eta = 0.7$ and $\beta_0 = 1.085$. Here the V-AV pair is produced inside the junction, 
unlike J vortices which only penetrate through the edges \cite{physc}. }
\end{figure}

\begin{figure} 
\includegraphics[width=\columnwidth]{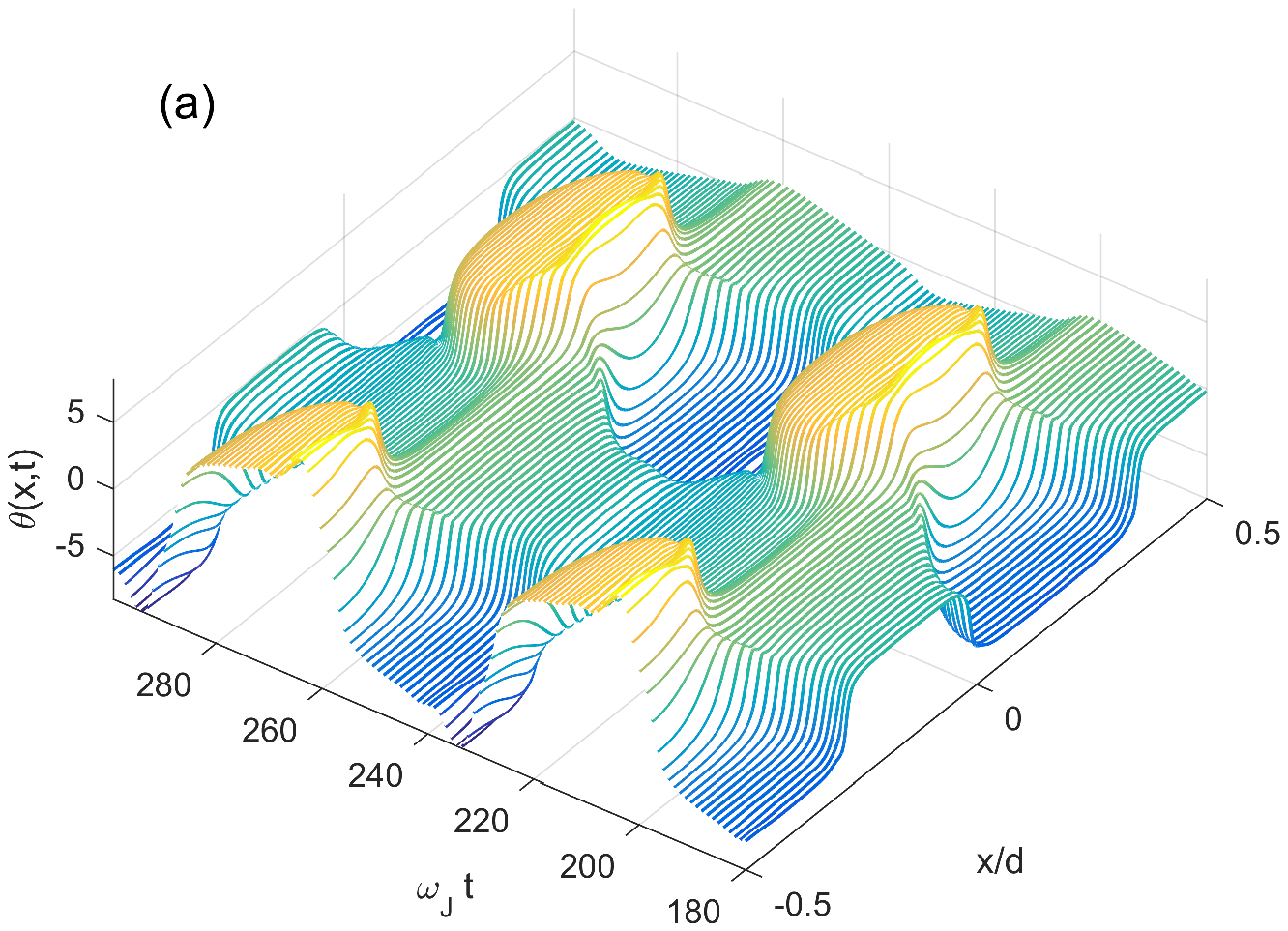}
\\
\includegraphics[width=\columnwidth]{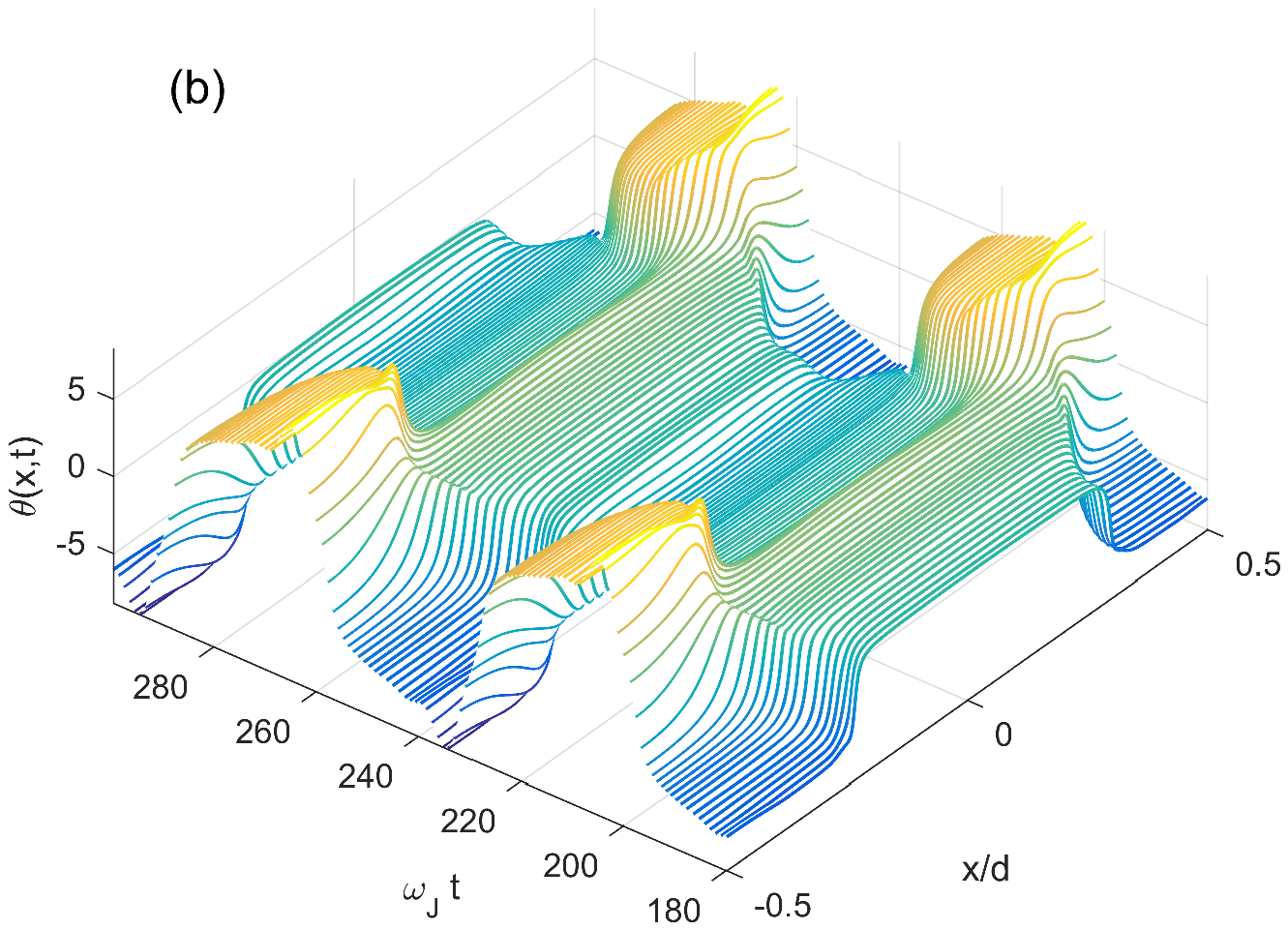}
\\
\includegraphics[width=\columnwidth]{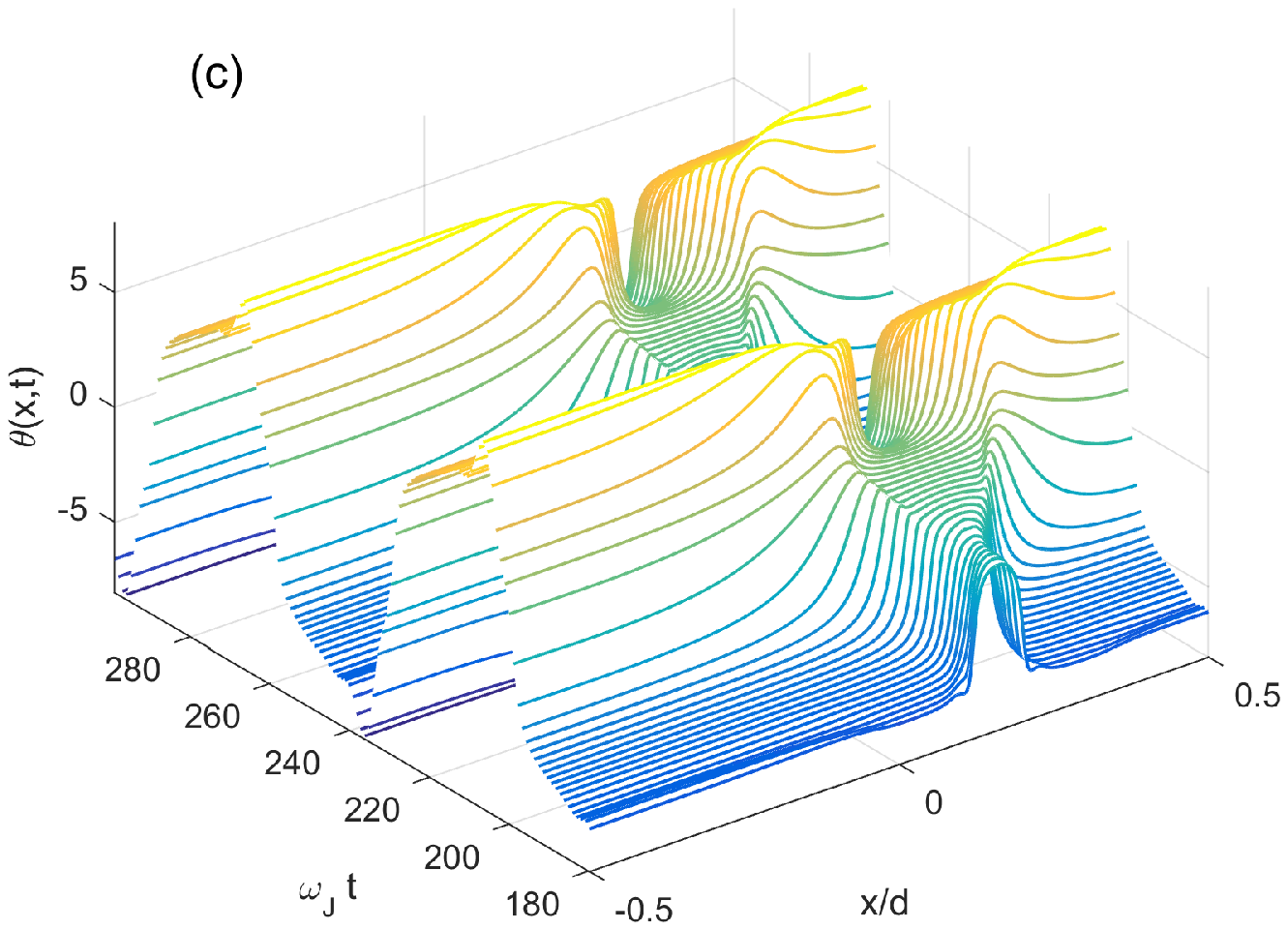}
\caption{\label{acins} Dynamic vortex patterns calculated at $\eta=0.7$ and: $\beta_0=1.092$ (a); $\beta_0 = 1.098$ (b); $\beta_0 = 1.149$ (c).}
\end{figure}

At smaller damping constants $0.3<\eta<0.7$ the radiation field produced by AJ vortices (see Fig. \ref{acrad}) makes their dynamic behavior rather different from that of J vortices described by the sine-Gordon equation \cite{physc}. As an illustration, we discuss here the underlying dynamics of vortices behind the behavior of $\bar{P}(\beta_0)$ at $\eta=0.7$ shown in Fig. \ref{acprofiles}.  Here the first jump on the ascending branch of $\bar{P}(\beta_0)$ at $\beta_0= 1.068$ results from penetration of a vortex in the positive ac cycle and an antivortex in the negative ac cycle. Figure \ref{acvav} shows the simulated dynamics of an AJ vortex which enters the junction during a positive ac cycle, stops midway when $\beta(t)$ changes sign, turns around and accelerates toward the left edge due to the combined effect of the Lorentz force and the attraction to the edge of the junction.  As the vortex approaches the edge, it leaves behind a radiation wake which eventually produces a V-AV pair. Then two vortices exit from the left edge of the junction while the remaining antivortex moves to the right, repeating the path of the vortex during the positive ac cycle.  At $\beta_0 > 1.092$, a radiation wake caused by oscillating vortices can produce a V-AV pair deep inside the junction \cite{supp}, resulting in the second jump in $\bar{P}(\beta_0)$ at $\beta_0=1.092$ in Fig. \ref{acprofiles}. In the range of $1.092<\beta_0 < 1.098$ a vortex/antivortex periodically entering and exiting from the left edge of the junction coexists with an oscillating V-AV pair, as shown in Fig. \ref{acins} (a). At $\beta_0>1.098$, the amplitude of relative V-AV oscillations increases and one component of the pair exits from the right edge. As a result, only one vortex and one antivortex remain in the junction, and the power $\bar{P}(\beta_0)$ drops, as shown in Figs. \ref{acprofiles} and \ref{acins} (b).  As $\beta_0$ increased further, vortices and antivortices penetrating from the opposite edges become closer to each other and eventually merge, evolving into the phase slip state as shown in Figs. \ref{acins} (b) and \ref{acins} (c).  On the descending branch of $\bar{P}(\beta_0)$ the phase slip state goes back to counter-oscillating vortex and antivortex penetrating from the opposite edges from $\beta_0 < 1.132$ down to $\beta _0= 0.875$ at which no vortices exist in the junction.

At $\eta=0.4$ the first big jump on the ascending branch of $\bar{P}(\beta_0)$ shown in Fig. \ref{acprofiles} occurs at $\beta_0 =1.034$ as two radiating vortices penetrates the junction during the positive ac cycle, stop midway and return during the negative ac cycle. Similar to the case of $\eta = 0.7$, each of these two vortices produce a V-AV pair, then all vortices exit and two antivortices remain. In turn, these antivortices repeat the same process during the negative ac cycle. As $\beta_0$ increases vortices penetrate deeper into the junction until the motion of the vortex pair becomes ballistic and $\bar{P}(\beta_0)$ drops at $\beta_0= 1.068$. At higher current signs of chaotic dynamics of oscillating vortices coexisting with ballistic vortices appear. In this region of $\beta_0\simeq 1.2-2$ simulations of Eq. (\ref{main}) become very time consuming and do not converge to an apparent time-periodic solution. Yet as $\beta_0$ further increases, the phase slip state eventually takes over so that $\bar{P}(\beta_0)$ becomes close to $\bar{P}(\beta_0)$ of a point junction and turns into a quadratic dependence at larger ac amplitudes. On the descending branch of $\bar{P}(\beta_0)$, counter-oscillating vortex and antivortex remain in the junction all the way to $\beta_0 = 0.89$. At lower currents a step in $\bar{P}(\beta_0)$ at $\beta_0 = 0.8$ occurs as only one vortex remains in the junction during positive ac cycle followed by one antivortex  during negative ac cycle, until neither of them can exist in the junction at $\beta_0<0.57$.

\begin{figure} 
\includegraphics[width=\columnwidth]{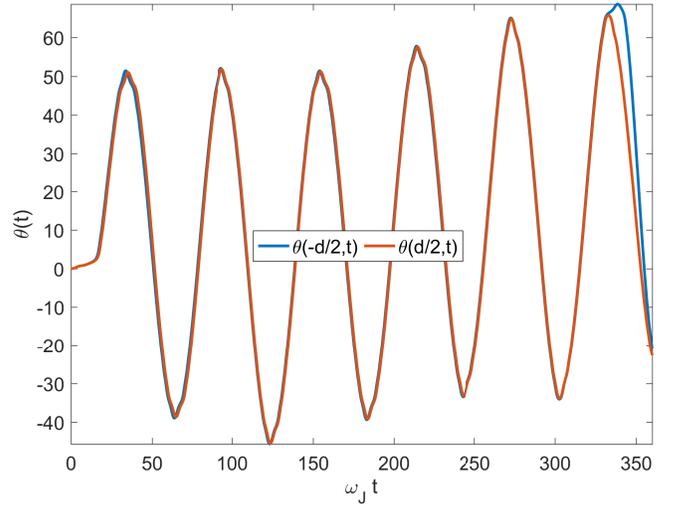}
\caption{\label{acphtm} Dynamics of $\theta(-d/2,t)$ and $\theta(d/2,t)$ at the edges at $\eta=0.2$ and $\beta_0=1.1$. Here $\theta(x,t)$ remains nearly uniform along junction, indicating a phase slip behavior.}
\end{figure}

Our simulations of Eq. (\ref{main}) at $\eta < 0.3$ have shown that the vortex starts producing a cascade of V-AV pairs right after it enters the junction which thus switches into a stochastic phase slip state coexisting with intermittent vortices and antivortices even in high currents. For instance, Fig. \ref{acphtm}, which shows $\theta(\pm d/2,t)$ at the edges, illustrates the dominance of phase slip state in junction for most of the time and the appearance of a vortex at $t\simeq 320$. Similar results were observed for the case of a point defect at the edge of the junction under ac current (more simulations can be found in Ref. \onlinecite{supp}).  

\section{Discussion}
\label{sec:disc}

In this paper we addressed nonlinear dynamics of vortices driven by strong dc and ac currents in Josephson junctions for which nonlocality of Josephson electrodynamics is essential.  Behavior of AJ  vortices in such junctions turns out to be different from either J or A vortices. Our numerical simulations and analytical results show that as current increases, moving AJ vortex structures evolve into a dynamic phase slip state similar to that of a point junction. This vortex-to-phase slip transition caused by the Josephson nonlocality occurs even in junctions much longer than the static AJ core length $l_0$, but the mechanisms of this transition are markedly different in overdamped and underdamped junctions.  In overdamped junctions the vortex-to-phase slip transition occurs because the length of the vortex core increases strongly as current increases, so that the vortex solutions disappear as the length of the vortex becomes of the order of the length of the junction. This conclusion follows from our exact solution for a driven AJ vortex at $\eta\gg 1$ and numerical simulations of Eq. (\ref{main}).     

In underdamped junctions the vortex-to-phase slip transition results from radiation of vortices which produce strong Cherenkov wakes and bremsstrahlung caused by interaction of vortices with the junction edges and other vortices.  These effects trigger generation of V-AV pairs inside the junction which become more pronounced as the damping constant $\eta$ decreases.  At $\eta < 0.3$ our simulations show that the vortex penetrating from the edge of the junction produces a cascade of expanding V-AV pairs driving the entire junction into the phase slip state.  In this case the $V-I$ curves become hysteretic, vortices emerge from the phase slip state as the current is decreased on the return branch of $V(I)$.  Dynamics of vortices driven by ac currents appears stochastic at small $\eta$ and $\beta_0\sim 1$, while the phase slip behavior is still dominant at ac amplitudes $\beta_0\gg 1$.  

Our calculations of $V-I$ characteristics and the power $P(\beta_0)$ dissipated by moving vortices show that $V(\beta_0)$ and $P(\beta_0)$ can be complicated functions of the amplitude $\beta_0$ of dc or ac current, and have regions with negative differential resistance $dV/dI$ and jump-wise hysteretic transitions. This situation is particularly relevant to underdamped junctions and grain boundaries at low temperatures in such materials like Nb$_3$Sn, iron-based superconductors and cuprates in which grain boundaries behave as planar weak links \cite{HM,D}.  In this case vortices moving along networks of grain boundaries of these polycrystalline materials can significantly contribute to the flux flow resistance and power dissipated under dc or ac currents, resulting in new mechanisms of nonlinearity of electromagnetic response associated with the dynamics of AJ vortices. These effects are essential for the understanding of the nonlinear residual surface resistance in polycrystalline resonator cavities and thin film multilayer screens under strong RF electromagnetic field.   

Proliferation of V-AV pairs caused by moving vortices can be essential for weak link superconducting structures in which the dynamic vortex instabilities  
can result in hysteretic jumps on the $V-I$ curves which appear similar to those produced by overheating \cite{KL}. However, neither the dynamic  phase slip transition nor generation of V-AV pairs
are influenced by cooling conditions, although heating can mask these effects at $\eta\sim 1$. Heating is most pronounced in overdamped junctions in which 
radiation is suppressed, while the generation of V-AV pairs is characteristic of underdamped junctions. Yet the jumps of the $V-I$ curves caused by penetration of vortex bundles in 
underdamped junctions can result in local heating which, in turn, can trigger thermal instabilities similar to those for A vortices under strong ac fields \cite{gigi}. 

The effects addressed in this work do not require special junctions with $J_c\sim J_d$. Indeed, the Cherenkov instability caused by weak NJE effects occurs 
even in a planar weak link with $\lambda_J = 10\lambda$ shown in Fig. \ref{scireppro}, whereas in thin film edge junctions the nonlocality becomes essential at much
lower $J_c$. Interaction of J or AJ vortices with pinned A vortices in electrodes can bring about additional
mechanisms of splitting instability of vortices. For instance, radiation by AJ
vortices can be enhanced as they move in a periodic magnetic potential of A vortices
along grain boundaries \cite{ajgb1,gc}, whereas A vortices trapped perpendicular to the 
junction can result in generation of V-AV pairs in the presence of the applied
electric current\cite{pet}. The result of this work may also pertain to the transition of A vortices driven by strong currents into chains of weakly coupled J vortices 
or phase slips in wide thin films \cite{ps1,ps2,ps3,ps4,ps5,ps6,ps7}.  In this case vortices moving along a self-induced channel of reduced order parameter behave 
as overdamped AJ vortices considered here.  As the current increases the AJ vortices further elongate along the flux channel and move faster, so we may expect  
a transition from the AJ vortices to a phase slip state above a threshold current in a film strip, similar to that for a Josephson junction of finite length.

\section*{Acknowledgments}

This work was supported by the US Department of Energy under Grant No. DE-SC0010081.

\appendix
\section{Derivation of Eq. (\ref{main})}
\label{app:A}
Equation (\ref{gxy}) gives $g(x,0,t)$ on the junction:
\begin{gather}
g(x,0,t)= 
\nonumber \\
- Jx -\frac{c\phi_0}{16\pi^3\lambda^2}\int_{-d/2}^{d/2}\ln\left|\frac{\cos\frac{\pi}{2d}(x+u)}{\sin\frac{\pi}{2d}(x-u)}\right|\theta'(u,t)du.
\label{wg}
\end{gather}
Using Eq. (\ref{wg}) we calculate $J_y(x,0,t)=-\partial_x g(x,0,t)$ and integrate the result by parts:
\begin{gather}
J_y(x,0,t) = J -
\nonumber \\
\frac{c\phi_0}{32\pi^2\lambda^2d}\int_{-d/2}^{d/2}\left[\cot\frac{\pi}{2d}(x-u)+\tan\frac{\pi}{2d}(x+u)\right]\theta'(u)du
\nonumber \\
=J+\frac{c\phi_0}{16\pi^3\lambda^2}\left[\ln\left|\frac{\sin\pi x/d-\sin\pi u/d}{2}\right|\theta'(u)\right]_{-d/2}^{d/2}+
\nonumber \\
 \frac{c\phi_0}{16\pi^3\lambda^2}\int_{-d/2}^{d/2}\ln\left|\frac{2}{\sin\pi x/d-\sin\pi u/d}\right|\theta''(u,t)du. 
\end{gather}
Here $\theta'(\pm d/2)=0$ because $J_x(\pm d/2)=0$ at the ends of the junction. Equating $J_y$ to the sum of Josephson, resistive, 
and displacement current densities,  we obtain: 
\begin{gather}
\ddot{\theta}+\eta\dot{\theta}+\sin\theta - \beta=
\nonumber \\
\left(\frac{\lambda_J^2}{\pi\lambda}\right)\int_{-d/2}^{d/2}\ln\left|\frac{2}{\sin\pi x/d-\sin\pi u/d}\right|\theta''(u)du, 
\label{nje}
\end{gather} 
where $\beta =J/J_c$. Equation (\ref{nje}) in which $x$ and $u$ are expressed in units of $d$, and $\epsilon=\lambda_J^2/\pi \lambda d$ reduces to Eq. (\ref{main}) which was used in our simulations.

Now we turn to $\theta(x,t)$ after the transition from the vortex to a phase slip state in which 
\begin{equation}
\theta(x,t)=\theta_0(t)+\theta_s(x),
\label{ps}
\end{equation}
where $\theta_0(t)$ satisfies Eq. (\ref{t0t}) for a point JJ.
The small stationary term $\theta_s(x)$ results from the slight inhomogeneity of $\beta(x)=(1-kx)\beta_0$ due to weak screening. 
Substituting Eq. (\ref{ps}) into Eq. (\ref{nje}) we see that the term $\sin\theta\simeq \sin\theta_0(t)+\theta_s\cos\theta_0(t)$ oscillates rapidly so  $\theta_s(x)\cos\theta_0(t)$
yields a small dynamic correction $\delta\theta(x,t)$ to $\theta(x,t)$ which is negligible at large $\beta$ and small $k$ we are interested in. The static $\theta_s$ can be calculated from 
Eq. (\ref{lond}) with $H=0$ by presenting $\theta_s(x)$ in the form 
which satisfies the boundary conditions $\theta_s'(\pm 1/2)=0$:
\begin{equation}
\theta_s(x)=\sum_{n=0}^\infty \theta_n\sin q_nx,
\label{ext}
\end{equation}
where $q_n=\pi(2n+1)/d$. Solution of Eq. (\ref{lond}) is then
\begin{equation}
g(x,y)=-\frac{c\phi_0}{16\pi^2\lambda^2}\sum_{n=0}^\infty\theta_ne^{-q_n|y|}\cos q_nx
\label{gg}
\end{equation}
From $kJx/d=-\partial_xg(x,0)$, it follows that 
\begin{equation}
Jk\frac{x}{d}=-\frac{c\phi_0}{16\pi^2\lambda^2}\sum_{n=0}^\infty\theta_n q_n\sin q_nx.
\label{b1}
\end{equation}
Multiplying both sides of Eq. (\ref{b1}) by $\sin q_{m}x$ and integrating from $-d/2$ to $d/2$ yields:
\begin{equation}
\frac{2Jk(-1)^n}{dq_n^2}=-\frac{c\phi_0d q_n\theta_n}{32\pi^2\lambda^2}.
\label{b2}
\end{equation}
Hence, $\theta_n=-4\beta_0 k(-1)^n/\pi^4\epsilon (2n+1)^3$, and
\begin{equation}
\theta_s(x)=-\frac{4\beta_0 k}{\pi^4\epsilon}\sum_{n=0}^\infty\frac{(-1)^n}{(2n+1)^3}\sin\frac{\pi x}{d}(2n+1),
\label{b3}
\end{equation} 
where $\epsilon = \lambda_J^2/\pi\lambda d$ and $\beta_0 = J/J_c$. 

\section{Exact solution for AJ vortex.}
\label{app:B}

The phase difference $\theta(x)=w_2(x,0)-w_1(x,0)$ is calculated using Eqs. (\ref{w1}) and (\ref{w2}), where the imaginary parts of $w_1$ and $w_2$ cancel out at $y=0$ because of continuity of $g_1(x,0)=g_2(x,0)$:
\begin{gather}
i\theta=\ln\frac{\sin\frac{\pi}{2}(x+u-il)}{\sin\frac{\pi}{2}(x-u-il)}-\ln\frac{\sin\frac{\pi}{2}(x+u+il)}{\sin\frac{\pi}{2}(x-u+il)} +i\chi
\nonumber \\
=\ln\frac{\cos\pi(u-il)-\cos\pi x}{\cos\pi(u+il)-\cos\pi x}+i\chi.
\label{i1}
\end{gather}
The time derivative of Eq. (\ref{i1}) yields 
\begin{gather}
\dot{\theta}-\dot{\chi}=
\frac{\pi\dot{u}}{D}(\sinh2\pi l-2\cos\pi x\cos\pi u\sinh\pi l)
\nonumber \\
+\frac{\pi\dot{l}}{D}(\sin 2\pi u-2\cos\pi x\sin\pi u\cosh\pi l),
\label{i2} 
\end{gather}
where the overdot denotes differentiation with respect to the dimensionless time $t/\tau$, and
\begin{equation}
D=\cos^2\!\pi u+\sin^2\!\pi l-2\cos\pi x\cos\pi u\cosh\pi l+\cos^2\!\pi x.
\label{D}
\end{equation}
Using Eqs. (\ref{jj}) and (\ref{i1}), we calculate:
\begin{gather}
\sin(\theta-\chi)=
\frac{2}{D}\sin\pi u\sinh\pi l(\cos\pi u\cosh\pi l-\cos\pi x),
\label{i3} \\
j_y=\frac{2\pi^2\epsilon}{D}\sin\pi u(\cos\pi u-\cos\pi x\cosh\pi l)+\beta.
\label{i4} 
\end{gather} 
Eq. (\ref{i1}) is an exact solution for AJ vortex, provided that the parameters $\chi$, $u$ and $l$ are such that 
the following boundary condition at the junction is satisfied:
\begin{equation}
\dot{\theta}+\sin\theta =  j_y.
\label{bca}
\end{equation}
Eqs. (\ref{i2})-(\ref{i4}) show that $\dot{\theta}-\dot{\chi}$, $\sin\theta$ and $j_y$ have the common denominator $D$ which is a quadratic polynomial in $\cos\pi x$, 
and their numerators are linear polynomials in $\cos\pi x$. Thus, Eq. (\ref{bca}) can be reduced to $A\cos^2\pi x+B\cos\pi x+C=0$, where $A$, $B$ and $C$ are independent of $x$.
Equating separately $A$, $B$ and $C$ to zero, we obtain that Eq. (\ref{i1}) is indeed the exact solution for AJ vortex in which $\chi(t)$, $u(t)$ and $l(t)$ satisfy the following equations 
\begin{gather}
\dot{\chi}+\sin\chi=\beta(t),
\label{x1} \\
\dot{l}=-\sin\pi u\sinh\pi l \times 
\nonumber \\
\frac{(\sin\pi u\cosh\pi l\cos\chi+\cos\pi u\sinh\pi l\sin\chi)}{\pi(\sinh^2\pi l+\sin^2\pi u)} +\pi\epsilon,
\label{x2} \\
\dot{u}=\sin\pi u\sinh\pi l \times 
\nonumber \\
\frac{(\sin\pi u\cosh\pi l\sin\chi-\cos\pi u\sinh\pi l\cos\chi)}{\pi(\sinh^2\pi l+\sin^2\pi u)}.
\label{x3} 
\end{gather}
Equations (\ref{x2}) and (\ref{x3}) are real and imaginary parts of a single complex differential equation (\ref{e2}).

For a vortex at the edge of a junction ($\pi u\ll d$ and $\pi l\ll d$), Eqs. (\ref{x2}) and (\ref{x3}) reduce to \cite{silin}:
\begin{gather}
\tau\partial_t u=\frac{ul}{u^2+l^2}(u\sin\chi-l\cos\chi),
\label{s1} \\
\tau\partial_t l=-\frac{ul}{u^2+l^2}(u\cos\chi+l\sin\chi)+l_0
\label{s2}
\end{gather} 
For a vortex in the middle of the junction $(u=1/2)$ at $\beta=0$, Eqs. (\ref{x2}) and (\ref{x3}) become:
\begin{gather}
\pi\dot{l}=-\tanh\pi l+\pi^2\epsilon, 
\label{mid1} \\
\dot{u}=0.
\label{mid2}
\end{gather}
The vortex at $u=1/2$ is in unstable equilibrium. To show that, we linearize Eqs. (\ref{x2}) and (\ref{x3}) with respect to
small perturbations $\delta l(t)$ and $\delta u(t)$ around the equilibrium values of $l$ and $u$ and obtain the following equations:
\begin{gather}
\delta\dot{l}=-(1-\pi^4\epsilon^2)\delta l,
\label{dl} \\
\delta\dot{u}=\pi^4\epsilon^2\delta u.
\label{du}
\end{gather}
Here we used the equilibrium relations $u=1/2$, $\tanh (\pi l)=\pi^2\epsilon$, and $\mbox{sech}^2 (\pi l) = 1-\pi^4\epsilon^2$. 
Eqs. (\ref{dl}) and (\ref{du}) describe two decoupled relaxation modes:
\begin{gather}
\delta l(t)=\delta l(0)e^{t\gamma_l}, \qquad \gamma_l=-1+\pi^4\epsilon^2,
\label{ii1} \\
\delta u(t)=\delta u(0)e^{t\gamma_u}, \qquad \gamma_u=\pi^4\epsilon^2,
\label{ii2}
\end{gather}
where $\gamma_l$ and $\gamma_u$ are decrements of perturbations of the core length and position, respectively. Here $\gamma _l $ is negative if $\pi^2\epsilon<1$ so the  
vortex breathing mode decays exponentially with the time constant $t_l=\tau\gamma_l^{-1}$ diverging at the phase slip transition $\epsilon=\pi^{-2}$.  
However, small displacements of the vortex increase exponentially with the time constant $t_u=\tau/\pi^4\epsilon^2$.  As the length of the junction decreases, $t_u\propto d^2$ decreases 
and approaches $\tau$ at $\epsilon=\pi^{-2}$.

\section{Numerical method}
\label{app:C}

We have developed an efficient MATLAB numerical code to solve the integro-differential equation
(\ref{main}) using the method of lines\cite{mdln}. By discretizing the integral term in Eq. (\ref{main}) we reduced it to a set of coupled nonlinear ordinary differential equations in time which were solved by the multistep, variable order Adams-Bashforth-Moulton method\cite{mdabm}. The absolute
and relative error tolerances were kept below $10^{-6}$. We have also checked our numerical results using a slower iterative
method to ensure the validity of results.  The steady state phase distribution $\theta(x-vt)$ at a given $\beta$ was computed by solving Eq. (\ref{main}) with zero initial conditions. The code then runs until a periodic solution - if applicable - is attained.

\end{document}